\documentclass[pdflatex,sn-basic]{sn-jnl}

\usepackage{braket}
\usepackage{graphicx}
\usepackage[all]{hypcap}
\usepackage{xcolor}
\usepackage{array}
\usepackage{float}
\usepackage{mathtools}
\usepackage{multirow}
\usepackage{colortbl}
\usepackage{comment}
\usepackage{algorithm}
\usepackage{algorithmicx}
\usepackage{algpseudocode}
\usepackage{amsmath}
\usepackage{amsfonts}
\usepackage{amssymb}
\usepackage[T1]{fontenc}
\usepackage{booktabs}
\usepackage{placeins} 

\theoremstyle{thmstyleone}%
\theoremstyle{thmstyletwo}%

\theoremstyle{thmstylethree}%

\begin{document}
\title[QShield: Adversarial Robustness via Quantum Circuits]{QShield: Securing Neural Networks Against Adversarial Attacks using Quantum Circuits}
%
%
\author*[1]{\fnm{Navid} \sur{Azimi}}\email{navid.azimi@emory.edu}
\author[1]{\fnm{Aditya} \sur{Prakash}}\email{aditya.prakash@alumni.emory.edu}
\author[2]{\fnm{Yao} \sur{Wang}}\email{yao.wang@emory.edu}
\author[1]{\fnm{Li} \sur{Xiong}}\email{lxiong@emory.edu}

\affil*[1]{\orgdiv{Department of Computer Science}, \orgname{Emory University}, \orgaddress{\street{201 Dowman Drive}, \city{Atlanta}, \postcode{30322}, \state{GA}, \country{USA}}}

\affil[2]{\orgdiv{Department of Physics}, \orgname{Emory University}, \orgaddress{\street{201 Dowman Drive}, \city{Atlanta}, \postcode{30322}, \state{GA}, \country{USA}}}
\abstract{Deep neural networks remain highly vulnerable to adversarial perturbations, limiting their reliability in security- and safety-critical applications. To address this challenge, we introduce QShield, a modular hybrid quantum--classical neural network (HQCNN) architecture designed to enhance the adversarial robustness of classical deep learning models. QShield integrates a convolutional neural network (CNN) backbone for feature extraction with a quantum processing module that encodes classical features into quantum states, applies structured entanglement under realistic noise models, and generates a hybrid prediction via a dynamically weighted multilayer perceptron. We systematically evaluate classical and hybrid models on the MNIST, OrganAMNIST, and CIFAR-10 datasets using comprehensive robustness and efficiency metrics.
Our results demonstrate that while classical models are highly susceptible to attacks, the proposed entangled hybrid models maintain high predictive accuracy and substantially reduce attack success rates across a diverse suite of adversarial attacks.
Furthermore, QShield significantly increases the computational cost of generating adversarial examples, introducing an additional layer of defense.
These findings indicate that the proposed modular hybrid architecture achieves a practical balance between predictive accuracy and adversarial robustness, positioning it as a promising approach for secure and reliable machine learning in sensitive applications.}


\maketitle


\section{Introduction}
Neural networks, particularly convolutional neural networks (CNNs), are well-known to be vulnerable to adversarial examples, inputs that have been subtly perturbed to cause a model to misclassify with high confidence \citep{west2023benchmarking}. Over the past several years, various adversarial attack methods have exposed these vulnerabilities \citep{goodfellow2014explaining,kurakin2018adversarial,madry2017towards,carlini2017towards,moosavi2016deepfool,andriushchenko2020square,liang2022adversarial}. To counter these attacks, numerous defense techniques have been proposed, including adversarial training, preprocessing techniques, randomization, and detection mechanisms \citep{wang2022adversarial}. Despite these efforts, achieving reliable robustness remains challenging, as many proposed defenses have been shown to fail under stronger or adaptive attacks \citep{athalye2018obfuscated,tramer2020adaptive}. Consequently, only a limited number of approaches provide consistent robustness guarantees, motivating the exploration of fundamentally different modeling paradigms. In this context, quantum-enhanced learning models have recently emerged as a promising direction that may offer new perspectives on adversarial robustness.

Hybrid quantum--classical neural networks integrate parameterized quantum circuits (PQCs) into classical deep learning models, leveraging the expressive power of quantum Hilbert spaces to enhance representation learning \citep{rizvi2025quantum}. Prior work has explored several hybrid designs, including quanvolutional layers, quantum pooling, and fully hybrid pipelines, demonstrating promising performance on vision tasks \citep{henderson2020quanvolutional,cong2019quantum,rizvi2025quantum}. Recent studies have also begun to investigate their adversarial robustness, indicating that architectural design choices significantly influence their resilience to adversarial attacks \citep{el2024advqunn}. While initial results indicate that quantum classifiers may inherit vulnerabilities similar to classical models \citep{liu2020vulnerability,lu2020quantum}, some evidence suggests potential robustness benefits under specific settings, including improved resistance to gradient-based attacks and noise-induced stabilization effects \citep{guan2021robustness,huang2023certified}. However, overall robustness behavior remains inconsistent and not yet fully understood, with architectural design choices playing a key role. These observations motivate further investigation of HQCNNs from an adversarial robustness perspective, particularly the relationship between quantum circuit design and model resilience.

\bmhead{Contributions}
In this paper, we introduce QShield, a modular hybrid quantum--classical neural network architecture that incorporates two key innovations. First, we design a parameterized quantum circuit supporting multiple entanglement patterns between qubits, including linear, star, and fully connected configurations. This design enables richer quantum correlations and more expressive feature representations under realistic noise conditions. Second, we propose an adaptive hybrid fusion mechanism that dynamically integrates quantum-derived predictions with classical predictions from the CNN backbone. Implemented via a lightweight multilayer perceptron (MLP), this module learns input-dependent weighting coefficients that control the relative contribution of each component on a per-sample basis, enabling the flexible integration of complementary representations. By adaptively balancing quantum and classical outputs, QShield combines the strong pattern-recognition capability of CNNs with the expressive feature space of quantum circuits, while mitigating their individual limitations. As a result, QShield aims to achieve a practical balance between accuracy, robustness, and adaptability in adversarial settings. The main contributions of this work are summarized as follows:

\begin{enumerate}
    \item \textbf{Modular Hybrid Quantum--Classical Architecture.} We present a hybrid quantum--classical architecture that integrates a parameterized quantum circuit with a CNN backbone. The overall design emphasizes modularity, allowing individual components, including the backbone network, feature extraction module, classical-to-quantum encoder, and underlying quantum circuit, to be independently replaced or upgraded. Within QShield, a classical-to-quantum feature encoder transforms classical feature representations into parameterized quantum rotations. The quantum circuit subsequently applies structured entangling operations together with an explicit noise-modeling layer, after which expectation values are measured and transformed into quantum probability outputs, which are subsequently processed by the hybrid fusion layer for final prediction.
    \item \textbf{Adaptive Hybrid Quantum--Classical Fusion.} We introduce a learnable fusion mechanism that adaptively integrates the outputs of the quantum and classical components for final prediction. This mechanism can be interpreted as a dynamic ensemble strategy, where the contribution of each component is adjusted based on its relative predictive strength. Similar to ensemble-based defenses in classical vision that improve robustness by aggregating multiple models \citep{strauss2017ensemble}, our approach extends this principle to a hybrid quantum--classical setting.
    \item \textbf{Comprehensive Adversarial Robustness Evaluation.} We conduct an extensive empirical evaluation across multiple datasets and model configurations, systematically analyzing the impact of different parameters and structures within the QShield architecture. The proposed models are evaluated under a diverse suite of adversarial attacks, including FGSM, PGD, APGD, VMI-FGSM, C\&W, DeepFool, One-Pixel, and Square Attack. We benchmark QShield against classical neural network baselines to quantify relative robustness and attack resilience.
\end{enumerate}


\section{Preliminaries}
\subsection{Quantum Computing}
A qubit serves as the fundamental unit of quantum information, analogous to a classical bit but governed by the principles of quantum mechanics. It constitutes a two-level quantum system with basis states denoted by $\ket{0}$ and $\ket{1}$. A general qubit state exists as a superposition $\ket{\psi} = \alpha \ket{0} + \beta \ket{1}$, where $\alpha$ and $\beta$ are complex amplitudes satisfying the normalization condition $|\alpha|^2 + |\beta|^2 = 1$ \citep{white2024magic}. Unlike a classical bit, which is restricted to a deterministic value of 0 or 1, a qubit can maintain a coherent superposition of both basis states until measurement occurs. Upon measurement, the state probabilistically collapses to one of the basis states, yielding outcome 0 with probability $|\alpha|^2$ or outcome 1 with probability $|\beta|^2$. This capacity for simultaneous state occupation implies that a system of $n$ qubits resides in a $2^n$-dimensional Hilbert space, formed by the tensor product of $n$ two-dimensional spaces. Consequently, an $n$-qubit system can collectively exist in a superposition of $2^n$ basis states, providing access to an exponentially large state space \citep{rieffel2000introduction}.

Quantum operations are executed via quantum gates, the quantum analogues of classical logic gates. Each gate corresponds to a unitary operator $U$ (a complex matrix satisfying $U U^\dagger = I$) that acts on one or more qubits. These unitary transformations are inherently reversible and preserve the normalization of the quantum state \citep{zaman2023survey}. Single-qubit gates perform rotations on the Bloch sphere, a geometric representation of a qubit's state, whereas multi-qubit gates create entanglement. Entanglement introduces nonclassical correlations between qubits, such that their joint state cannot be factored into independent single-qubit states. Key examples of common quantum gates include \citep{evans2024quick,du2025quantum}:

\begin{itemize}
    \item \textbf{Hadamard (H).} A single-qubit gate that transforms a basis state $\ket{0}$ or $\ket{1}$ into an equal superposition of both. Specifically, it maps $\ket{0} \mapsto (\ket{0}+\ket{1})/\sqrt{2} = \ket{+}$ and $\ket{1} \mapsto (\ket{0}-\ket{1})/\sqrt{2} = \ket{-}$. 
    \item \textbf{Pauli-X ($\boldsymbol{R_X}$).} A $180^\circ$ rotation about the X-axis (the quantum NOT gate) that flips the state such that $\ket{0} \leftrightarrow \ket{1}$. 
    \item \textbf{Pauli-Z ($\boldsymbol{R_Z}$).} A $180^\circ$ rotation about the Z-axis that introduces a phase flip, mapping $\ket{1}$ to $-\ket{1}$ while leaving $\ket{0}$ unchanged.
    \item \textbf{Pauli-Y ($\boldsymbol{R_Y}$).} A $180^\circ$ rotation about the Y-axis, combining a bit-flip and a phase flip, mapping $\ket{0}$ to $i\ket{1}$ and $\ket{1}$ to $-i\ket{0}$.
    \item \textbf{Controlled-NOT (CNOT).} A two-qubit entangling gate that flips the state of a target qubit if and only if the control qubit is in the state $\ket{1}$. For example, applying a CNOT gate to two qubits initially in the state $\ket{q_{\text{control}}, q_{\text{target}}} = \ket{1,0}$ yields $\ket{1,1}$, whereas the state $\ket{0,0}$ remains $\ket{0,0}$. The CNOT gate serves as a primary mechanism for creating entangled pairs.
\end{itemize}

A quantum circuit comprises a sequence of quantum gate operations applied to a collection of qubits, typically followed by measurement. The circuit generally initializes in the state $\ket{0}^{\otimes n}$ and terminates with projective measurements that yield a classical bit-string. Gates operating independently on distinct qubits are mathematically represented by the tensor product, such as $U \otimes V$ \citep{white2024magic}.

In this paper, we adopt a layer-wise formulation for a parameterized quantum circuit acting on \(n\) qubits \citep{wang2021noise}. The circuit is expressed as a product of \(L\) unitary layers:
\begin{equation}
U(\vec{\boldsymbol{\theta}})
=
U_{L}(\vec{\boldsymbol{\theta}_{L}}) \cdots U_{2}(\vec{\boldsymbol{\theta}_{2}}) U_{1}(\vec{\boldsymbol{\theta}_{1}}),
\end{equation}
where \(\vec{\boldsymbol{\theta}}=\{\vec{\boldsymbol{\theta}}_{\ell}\}_{\ell=1}^{L}\) denotes the set of rotation parameters. Although our implementation utilizes a variational circuit consisting of a single layer, we retain the layer index $\ell$ in the notation to maintain generality and accommodate future extensions to deeper circuits.

Each circuit layer consists of a parameterized single-qubit rotation block followed by a fixed multi-qubit entangling block (except in the no-entanglement configuration). The rotation block is defined as:
\begin{equation}
U_{\mathrm{rot}}(\vec{\boldsymbol{\theta}_{\ell}})
=
\bigotimes_{i=0}^{n-1}
\left(
R_{Z}\!\big(\theta^{(z)}_{\ell,i}\big)\,
R_{Y}\!\big(\theta^{(y)}_{\ell,i}\big)\,
R_{X}\!\big(\theta^{(x)}_{\ell,i}\big)
\right),
\end{equation}
where \(R_{\alpha}(\theta)=e^{-i\theta \sigma_{\alpha}/2}\) for \(\alpha\in\{X,Y,Z\}\) represents the single-qubit rotation gates, and \(\theta^{(x)}_{\ell,i}\), \(\theta^{(y)}_{\ell,i}\), and \(\theta^{(z)}_{\ell,i}\) specify the rotation angles for qubit \(i\) in layer \(\ell\).

The entangling block is characterized by a fixed unitary operator:
\begin{equation}
U_{\mathrm{ent}} \in \mathcal{U}(2^{n}),
\end{equation}
which is constructed as an ordered composition of two-qubit entangling gates acting on designated pairs of qubits according to a predefined connectivity structure. This connectivity pattern is treated as an architectural hyperparameter and remains fixed throughout training. Consequently, the resulting unitary transformation for layer \(\ell\) is given by:
\begin{equation}
U_{\ell}(\vec{\boldsymbol{\theta}_{\ell}})
=
U_{\mathrm{ent}}\,U_{\mathrm{rot}}(\boldsymbol{\vec{\theta}}_{\ell}).
\end{equation}

To accurately model the noise inherent in realistic quantum hardware, we utilize the density-matrix formalism \citep{nielsen2010quantum,weber2024constructing}. For a pure input state \(|\psi\rangle\), the corresponding density operator is defined as:
\begin{equation}
\rho_{\mathrm{in}} = |\psi\rangle\langle\psi|,
\end{equation}
whereas mixed states are described by general positive semidefinite density operators \(\rho_{\mathrm{in}}\) that satisfy the trace condition:
\begin{equation}
\mathrm{Tr}(\rho_{\mathrm{in}})=1.
\end{equation}
Here, the symbol $\mathrm{Tr}(\cdot)$ denotes the trace of a matrix, defined as the sum of its diagonal elements.

For a single layer, the ideal unitary evolution is succeeded by a noise channel $\mathcal{N}$ that acts locally on each qubit. The noisy evolution of layer \(\ell\) is expressed as:
\begin{equation}
\rho_{\ell}
=
\mathcal{N}^{(\ell)}_{\mathrm{global}}
\!\left(
U_{\ell}(\vec{\boldsymbol{\theta}_{\ell}})\,\rho_{\ell-1}\,
U_{\ell}^{\dagger}(\vec{\boldsymbol{\theta}_{\ell}})
\right),
\qquad
\mathcal{N}^{(\ell)}_{\mathrm{global}}
=
\bigotimes_{i=0}^{n-1}\mathcal{N}^{(\ell)}_{i},
\end{equation}
with the initial condition \(\rho_{0}=\rho_{\mathrm{in}}\). This factorized representation assumes local Markovian noise acting independently on each qubit. Specifically, we neglect spatially or temporally correlated errors, crosstalk, and explicit two-qubit (gate-dependent) noise terms, treating each $\mathcal{N}^{(\ell)}_i$ as a single-qubit completely positive trace-preserving (CPTP) channel \citep{ma2024tomography}. Furthermore, we apply the noise channel once per variational layer, that is, immediately following the ideal unitary $U_\ell$. While this layer-level approximation simplifies both mathematical analysis and empirical training, a more hardware-faithful model could alternatively insert noise channels after each primitive gate, particularly following two-qubit entangling operations.

By composing the noisy evolution across all \(L\) layers, the final output state of the circuit is obtained as:
\begin{equation}
\rho_{\mathrm{out}}
=
\mathcal{N}^{(L)}_{\mathrm{global}}
\circ
\mathcal{U}_{L}
\circ
\cdots
\circ
\mathcal{N}^{(1)}_{\mathrm{global}}
\circ
\mathcal{U}_{1}
\bigl(\rho_{\mathrm{in}}\bigr),
\end{equation}
where
\begin{equation}
\mathcal{U}_{\ell}(\cdot)
=
U_{\ell}(\vec{\boldsymbol{\theta}_{\ell}})(\cdot)U_{\ell}^{\dagger}(\vec{\boldsymbol{\theta}_{\ell}})
\end{equation}
denotes the unitary channel associated with layer \(\ell\).

Finally, a projective measurement in the computational basis is performed across all $n$ qubits to extract the resulting outcome probabilities.

\subsection{Evaluation Metrics}
To comprehensively evaluate the robustness and efficiency of the models, we employ three widely recognized evaluation metrics \citep{macas2024adversarial}: Original Detection Rate (ODR), Attack Success Rate (ASR), and Total Time Cost (TTC).

\begin{itemize}
    \item \textbf{Original Detection Rate.}  
    The ODR quantifies the classification accuracy of a model on clean, unperturbed test samples. Let $N_{\text{correct}}^{\text{clean}}$ denote the number of correctly classified samples in the clean test set, and $N_{\text{total}}$ the total number of test samples. The ODR is calculated as:
    \begin{equation}
    \text{ODR} = \frac{N_{\text{correct}}^{\text{clean}}}{N_{\text{total}}} \times 100\%
    \end{equation}
    \item \textbf{Attack Success Rate.}  
    The ASR measures the effectiveness of adversarial attacks by determining the proportion of originally correct predictions that are altered to incorrect classifications following perturbation. Let $N_{\text{misclassified}}^{\text{adv}}$ represent the number of adversarial samples that are misclassified, and $N_{\text{correct}}^{\text{clean}}$ the number of clean correctly classified samples. The ASR is defined as:
    \begin{equation}
    \text{ASR} = \frac{N_{\text{misclassified}}^{\text{adv}}}{N_{\text{correct}}^{\text{clean}}} \times 100\%
    \end{equation}
    \item \textbf{Total Time Cost.}  
    The TTC represents the total elapsed computational time required to train and evaluate a given model, thereby capturing the inherent trade-off between robustness and computational efficiency:
    \begin{equation}
    \text{TTC} = T_{\text{train}} + T_{\text{test}}
    \end{equation}
    where $T_{\text{train}}$ and $T_{\text{test}}$ denote the training and testing times, respectively.
\end{itemize}


\section{Related Work}
\subsection{Adversarial Example Attacks}
Following the initial discovery of adversarial examples by Szegedy et al. \citep{szegedy2013intriguing}, Goodfellow et al. \citep{goodfellow2014explaining} introduced one of the first computationally efficient attack methods: the Fast Gradient Sign Method (FGSM). FGSM perturbs an input in the direction of the gradient's sign to maximally increase the loss. Stronger iterative attacks subsequently emerged, such as the Basic Iterative Method (BIM) \citep{kurakin2018adversarial} and its extension, Projected Gradient Descent (PGD). These methods apply FGSM repeatedly using small step sizes, projecting the perturbed input back into the allowed perturbation norm ball after each iteration. Formalized by Madry et al. \citep{madry2017towards} as a universal first-order adversary, PGD is widely recognized as one of the most powerful gradient-based attacks.

Other white-box attacks have leveraged optimization techniques to discover minimal adversarial perturbations. The Carlini \& Wagner (C\&W) attack \citep{carlini2017towards} formulates the search for an adversarial example as a relaxed optimization problem, achieving high attack success rates with notably small $L_2$ distortions \citep{zuo2021exploiting}. Similarly, the DeepFool attack \citep{moosavi2016deepfool} proposes an efficient procedure to approximate the closest decision boundary distance for a given input. By iteratively linearizing the classifier and moving the input toward the nearest misclassification hyperplane, DeepFool generates perturbations that are often orders of magnitude smaller than those produced by FGSM. In addition to white-box attacks, black-box attacks have been developed to target models when gradient information is inaccessible. A prominent example is the Square Attack, a score-based black-box method that relies on a random localized search strategy \citep{andriushchenko2020square}. The Square Attack perturbs small square regions of the image with random noise, iteratively refining the location and magnitude of these perturbations.

\subsection{Adversarial Defenses}
To counter adversarial attacks, a wide array of defense techniques has been proposed, including adversarial training, input preprocessing, gradient masking or obfuscation, inference-time randomization, and adversarial-example detection \citep{wang2022adversarial}. Among these approaches, adversarial training, wherein models are trained on carefully crafted adversarial examples, has emerged as one of the most reliable methods by explicitly optimizing models for robustness \citep{athalye2018obfuscated,madry2017towards}. In particular, training with PGD-generated adversarial examples has been shown to significantly improve resistance against a broad spectrum of attacks \citep{madry2017towards}. Despite its effectiveness, adversarial training is computationally expensive and typically introduces a trade-off between robustness and clean accuracy.

Other defense strategies include input preprocessing techniques that attempt to remove adversarial perturbations, gradient masking or obfuscation methods that hide the gradient information exploited by attackers, randomization during inference to diminish attack reliability, and detection frameworks designed to identify and flag adversarial inputs \citep{qiao2026survey}. However, many early defenses that initially reported strong robustness were subsequently shown to fail under adaptive attacks. For instance, Athalye et al. \citep{athalye2018obfuscated} demonstrated that 7 out of 9 proposed defenses relying on gradient obfuscation could be defeated once attackers adapted their methodologies. Consequently, defenses must be rigorously evaluated against adaptive white-box attacks to meaningfully assess their security guarantees \citep{tramer2020adaptive}. To date, the most consistently effective defenses include adversarial training and a few provable methods, such as randomized smoothing, whereas many heuristic defenses have proven vulnerable to stronger attacks.

\subsection{Hybrid Quantum--Classical Neural Networks}
A foundational development in the domain of quantum-enhanced machine learning is the quanvolutional layer, pioneered by Henderson et al. \citep{henderson2020quanvolutional}, which employs quantum circuits as convolutional filters. This approach demonstrated improved test accuracy and faster training convergence compared to purely classical CNNs. Subsequent work extended this framework by introducing trainable quantum filters and quantum pooling operations, leading to further performance gains on benchmark vision datasets such as MNIST \citep{li2022image}. 

Architectural diversity in quantum--classical neural networks ranges from fully quantum analogues, such as the QCNN proposed by Cong et al. \citep{cong2019quantum} for quantum state classification, to complex hybrid pipelines. For instance, the QCQ-CNN architecture \citep{long2025hybrid} utilizes a sequence of fixed quantum filters for nonlinear feature extraction, classical layers for spatial processing, and trainable quantum classifiers to establish robust decision boundaries. Beyond performance gains, the hybrid models are noted for their parameter efficiency, often matching the accuracy of dense classical networks with significantly fewer effective parameters \citep{huang2023image,wang2024shallow}.


\section{Baseline Classical Neural Network Architectures}
To systematically investigate model robustness under diverse adversarial attacks, we evaluate two primary categories of models: (i) classical neural networks, comprising fully connected deep neural networks (DNNs) and convolutional neural networks (CNNs), and (ii) hybrid quantum--classical neural networks (HQCNNs), which feature four architectural variants distinguished by their quantum circuit entanglement patterns. The hybrid models are detailed in Section~\ref{sec:qshield}. This comparative methodology ensures a comprehensive evaluation of conventional architectures against quantum-enhanced neural networks.

\subsection{Fully Connected Deep Neural Networks}
\begin{figure}
    \centering
    \begin{minipage}{0.86\textwidth}
        \centering
        \includegraphics[width=\textwidth, trim=10 10 10 10, clip]{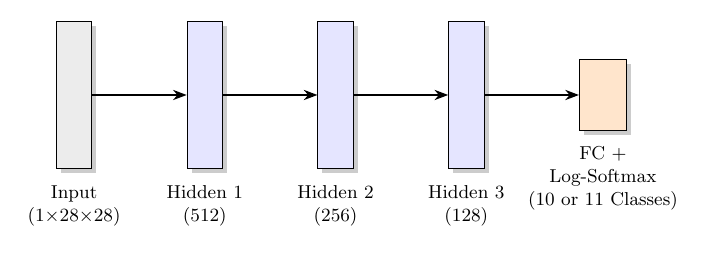}
        \vspace{0.5em}
        \small (a) DNN for MNIST (10 classes) and OrganAMNIST (11 classes)
    \end{minipage}
    \vspace{1.0em}
    \begin{minipage}{1\textwidth}
        \centering
        \includegraphics[width=\textwidth, trim=10 10 10 10, clip]{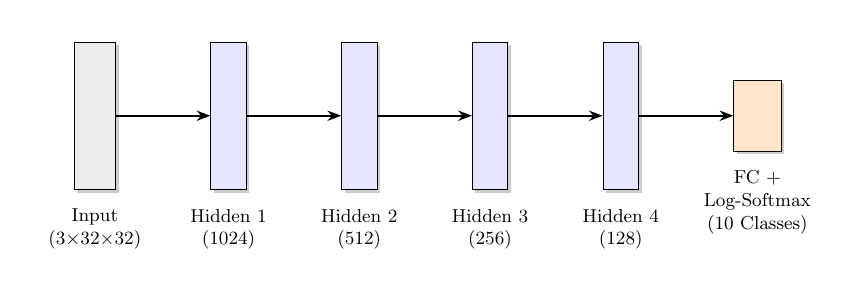}
        \vspace{0.5em}
        \small (b) DNN for CIFAR-10 (10 classes)
    \end{minipage}
    \caption{Architectural diagrams of the fully connected DNNs used for the MNIST, OrganAMNIST, and CIFAR-10 datasets.}
    \label{fig:dnn-architectures}
\end{figure}

We implement dataset-specific DNNs for the MNIST, OrganAMNIST, and CIFAR-10 datasets. Each hidden layer block adheres to a consistent design principle: stacked fully connected layers followed by batch normalization, ReLU activation functions, and dropout ($p=0.2$) for regularization. Model training is performed using the negative log-likelihood loss applied to log-softmax outputs. The resulting DNN architectures are illustrated in Fig.~\ref{fig:dnn-architectures} and summarized below:

\subsubsection{DNN-MNIST and DNN-OrganAMNIST}
Each $1 \times 28 \times 28$ grayscale image is flattened into a 784-dimensional vector. The network comprises three hidden layers with 512, 256, and 128 units, respectively. The final output layer maps these features to 10 classes for MNIST and 11 classes for OrganAMNIST.

\subsubsection{DNN-CIFAR10}
Each $3 \times 32 \times 32$ RGB image is flattened into a 3072-dimensional vector. The architecture incorporates four hidden layers containing 1024, 512, 256, and 128 units, respectively, followed by a 10-class output layer.

\subsection{Convolutional Neural Networks}
Our baseline CNN models are derived from a modified ResNet-18 \citep{he2016deep} backbone. Adaptations are introduced to accommodate varying input modalities (grayscale versus RGB) and to align with the specific output class structure of each dataset. Across all CNN variants, final predictions are generated by applying a log-softmax function to the logits produced by the adapted ResNet-18 backbone.

\subsubsection{Input Layer Adaptation}
For the grayscale datasets (MNIST and OrganAMNIST), the initial convolutional layer is replaced with a single-channel variant. The weights for this layer are initialized by averaging the pretrained RGB filters from an ImageNet-pretrained model. For the CIFAR-10 dataset, the original RGB input layer is retained.

\subsubsection{Output Layer Adaptation}
The final fully connected layer is modified to match the required number of target classes: 10 for MNIST and CIFAR-10, and 11 for OrganAMNIST.

\subsubsection{Dataset Variants}
\begin{figure}
    \centering
    \begin{minipage}{0.95\textwidth}
        \centering
        \includegraphics[width=\textwidth, trim=20 20 20 20, clip]{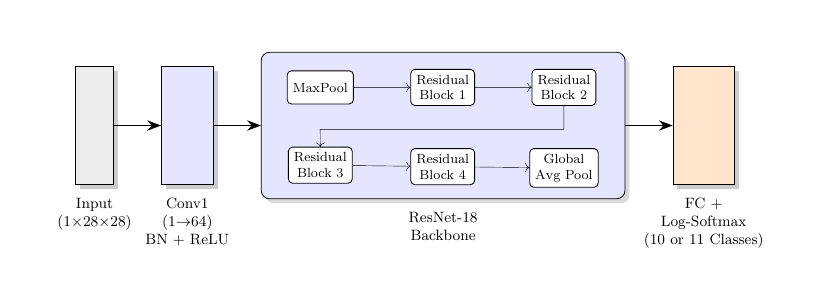}
        \vspace{0.5em}
        \small (a) CNN for MNIST (10 classes) and OrganAMNIST (11 classes)
    \end{minipage}
    \vspace{1.0em}
    \begin{minipage}{0.95\textwidth}
        \centering
        \includegraphics[width=\textwidth, trim=20 20 20 20, clip]{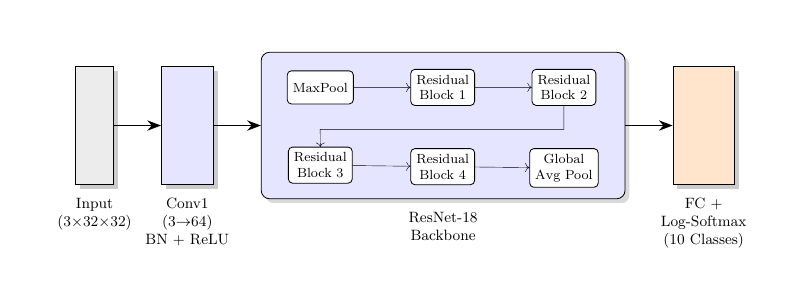}
        \vspace{0.5em}
        \small (b) CNN for CIFAR-10 (10 classes)
    \end{minipage}
    \caption{Architectural diagrams of the CNN models based on the ResNet-18 backbone, adapted for the MNIST, OrganAMNIST, and CIFAR-10 datasets.}
    \label{fig:cnn-architectures}
\end{figure}

The resulting dataset-specific CNN architectures are illustrated in Fig.~\ref{fig:cnn-architectures} and are defined as follows:
\begin{itemize}
    \item \textbf{CNN-MNIST:} Accepts a $1 \times 28 \times 28$ grayscale input and produces a 10-class output.
    \item \textbf{CNN-OrganAMNIST:} Accepts a $1 \times 28 \times 28$ grayscale input and produces an 11-class output.
    \item \textbf{CNN-CIFAR10:} Accepts a $3 \times 32 \times 32$ RGB input and produces a 10-class output.
\end{itemize}

\section{QShield Architecture}\label{sec:qshield}
\begin{figure}
  \centering
    \includegraphics[width=0.99\textwidth, trim=10 0 0 10, clip]{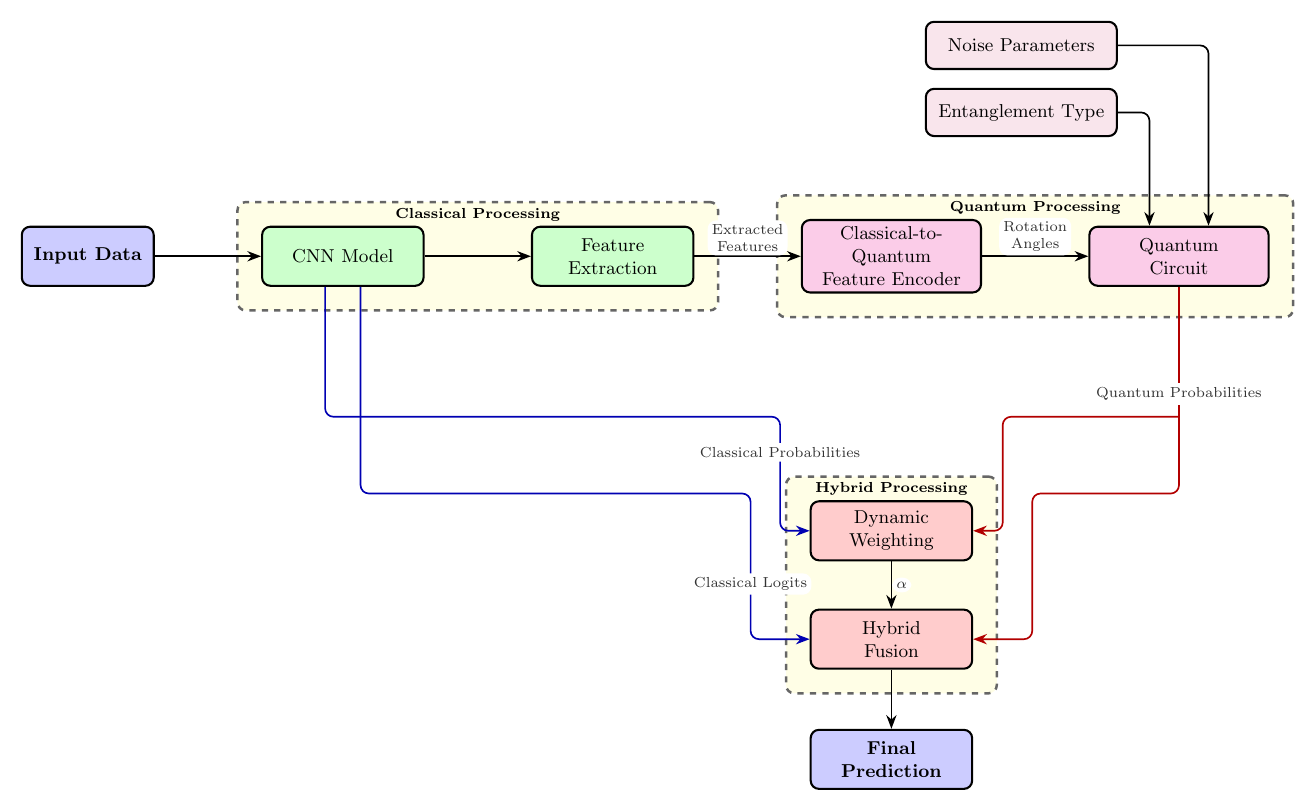}
    \caption{Schematic overview of the proposed QShield architecture. 
    The framework integrates classical CNN feature extraction, quantum processing via parameterized circuits (including entanglement and noise modeling), and a hybrid fusion stage governed by dynamic weighting.}
    \label{fig:qshield_architecture}
\end{figure}

We propose QShield, a hybrid quantum--classical neural network architecture designed to leverage the complementary strengths of both computational paradigms. A schematic overview of the QShield architecture is presented in Fig.~\ref{fig:qshield_architecture}. At a high level, QShield comprises four integrated components, each of which is described in detail in the subsequent subsections:

\begin{enumerate}
    \item \textbf{Feature Extraction and Quantum Encoding.} A classical CNN backbone extracts rich intermediate features from the input data, which are subsequently encoded into quantum states using a scheme that preserves the geometric and statistical properties of the data.  
    \item \textbf{Entanglement Patterns and Noise Modeling.} Four distinct entanglement patterns are implemented to explore their impact on model performance and robustness. A realistic quantum noise model is applied across all simulations to emulate hardware imperfections.  
    \item \textbf{Dynamic Fusion Coefficient.} A lightweight MLP adaptively computes a fusion coefficient $\alpha \in [0, 1]$, which regulates the relative contributions of the classical and quantum outputs.  
    \item \textbf{Hybrid Output Fusion.} The final prediction is generated by combining the classical and quantum outputs according to the learned fusion coefficient $\alpha$, effectively balancing conventional feature learning with quantum-enhanced representations. 
\end{enumerate}

\subsection{Feature Extraction and Feature Encoding}
Within the hybrid quantum--classical pipeline, feature extraction is performed by the CNN backbone, providing expressive intermediate representations that are subsequently encoded into quantum states. To facilitate this process, we adopt a flexible feature extraction module paired with an encoding mechanism designed to preserve the geometric and statistical characteristics of the data.

\subsubsection{CNN-Based Feature Extraction}
Algorithm \ref{alg:cnn-feature-extraction} details the procedure for extracting intermediate feature representations from the backbone CNN. The method leverages PyTorch forward hooks to non-invasively capture activations from a designated internal layer while preserving the standard forward computation of the model.

Given a CNN model $f_\theta$ and a batch of inputs $X$, the algorithm first identifies the target layer for feature extraction. If a specific layer name $L$ is provided, the corresponding submodule is selected. Otherwise, the algorithm automatically defaults to the final feature-producing layer, typically the last convolutional or linear layer preceding the classifier, ensuring compatibility across diverse CNN architectures.

A forward hook is then registered on the selected layer to record its output activations during the forward pass. The network is evaluated normally to produce the prediction $\hat{y}$, while the hook simultaneously captures the intermediate activation tensor $A$. Following the forward pass, the hook is detached to prevent side effects in subsequent evaluations.

Finally, the recorded activations are flattened along all non-batch dimensions to yield a feature matrix $F \in \mathbb{R}^{B \times D}$, where $B$ denotes the batch size and $D$ represents the resulting feature dimensionality. The algorithm returns both the network's prediction and the extracted feature representation, which are used for downstream processing.

\begin{algorithm}
\caption{CNN-Based Feature Extraction with Forward Hooks}
\label{alg:cnn-feature-extraction}
\begin{algorithmic}[1]
\Require CNN model $f_\theta$, input batch $X \in \mathbb{R}^{B \times \cdots}$, optional layer index/name $L$
\Ensure Prediction $\hat{y}$ and feature matrix $F \in \mathbb{R}^{B \times D}$
\Statex
\Statex \textbf{(1) Select Target Layer}
\State \textbf{if} $L$ is specified \textbf{then}
    \State \quad $m \gets$ submodule of $f_\theta$ identified by $L$
\State \textbf{else}
    \State \quad $m \gets$ last feature-extracting layer of $f_\theta$
\State \textbf{end if}
\Statex \textbf{(2) Register Forward Hook}
\State Initialize activation buffer $A \gets \varnothing$
\State Attach forward hook to $m$ that stores its output in $A$
\algstore{cnn-feature-extraction}
\end{algorithmic}
\end{algorithm}

\addtocounter{algorithm}{-1}

\begin{algorithm}
\caption{(continued)}
\begin{algorithmic}[1]
\algrestore{cnn-feature-extraction}
\Statex \textbf{(3) Forward Pass}
\State $\hat{y} \gets f_\theta(X)$
\State Let $A$ contain the captured activations
\Statex \textbf{(4) Hook Removal}
\State Detach forward hook from $m$
\Statex \textbf{(5) Feature Construction}
\State $F \gets \mathrm{Flatten}(A)$ over non-batch dimensions
\State $F \in \mathbb{R}^{B \times D}$
\State \Return $\hat{y}, F$
\end{algorithmic}
\end{algorithm}

\subsubsection{Feature Encoding for Quantum States}
Algorithm \ref{alg:feature-encoding} outlines the procedure for mapping classical feature representations into parameterized quantum rotations via an angle-encoding strategy. The goal of this encoding is to efficiently translate high-dimensional classical features into quantum gate parameters while respecting qubit resource constraints and ensuring numerical stability during training and inference.

Given a feature matrix $F \in \mathbb{R}^{B \times D}$ extracted from the classical backbone, the algorithm first applies batch-wise normalization by subtracting the mean and dividing by the standard deviation across the batch. This standardization ensures consistent feature scaling and stabilizes the subsequent nonlinear mapping to rotation angles.

To align the feature dimensionality with the quantum circuit requirements, the algorithm enforces a strict correspondence between the feature dimension and the number of available qubit rotation parameters. In our architecture, each qubit is parameterized by three independent rotation angles, fixing the target dimensionality at $3n$. When the feature dimension is smaller than this target, the algorithm projects the features into a higher-dimensional space using an orthogonally initialized linear transformation, preserving feature diversity and avoiding degenerate embeddings. Conversely, when the feature dimension exceeds $3n$, dimensionality reduction is applied: features exhibiting the highest variance are selected for moderately oversized inputs, whereas principal component analysis (PCA) is employed for very high-dimensional inputs to retain the most informative components.

The resulting feature matrix is reshaped into a three-dimensional tensor of shape $[B, n, 3]$, assigning a triplet of features to each qubit. These triplets are then mapped to rotation angles via a bounded nonlinear transformation using the hyperbolic tangent function scaled by $\pi$. This mapping guarantees that all rotation parameters are confined to the physically meaningful interval $(-\pi, \pi)$, ensuring compatibility with standard quantum gate implementations.

Finally, the algorithm outputs the angle tensor $\Theta \in \mathbb{R}^{B \times n \times 3}$, which is subsequently used to parameterize the corresponding rotation operator $U_{\mathrm{rot}}$ within the quantum circuit.

\begin{algorithm}
\caption{Feature Encoding for Quantum States via Angle Encoding}
\label{alg:feature-encoding}
\begin{algorithmic}[1]
\Require Feature matrix $F \in \mathbb{R}^{B \times D}$, number of qubits $n$
\Ensure Angle tensor $\Theta \in \mathbb{R}^{B \times n \times 3}$ for corresponding rotation operator $U_{\mathrm{rot}}$
\Statex
\Statex \textbf{(1) Normalization}
\State Compute batch-wise mean $\mu$ and standard deviation $\sigma$ over feature dimension
\State $F \gets (F - \mu) / \sigma$  \Comment{Standardize each feature for numerical stability}
\Statex \textbf{(2) Dimensionality Matching}
\If{$D < 3n$}
    \State Initialize linear map $W \in \mathbb{R}^{D \times 3n}$ with orthogonal columns (e.g., QR-based)
    \State $F \gets F W$  \Comment{Project to higher dimension while preserving feature diversity}
\ElsIf{$D > 3n$}
    \If{$D$ is moderately larger than $3n$}
        \State Select the $3n$ features with highest variance across the batch
        \State $F \gets \text{VarianceSelect}(F, 3n)$
    \Else
        \State Apply PCA to $F$ and retain top $3n$ principal components
        \State $F \gets \text{PCA}(F, 3n)$
    \EndIf
\EndIf
\State Now $F \in \mathbb{R}^{B \times 3n}$
\Statex \textbf{(3) Reshaping}
\State Reshape $F$ to $\tilde{F} \in \mathbb{R}^{B \times n \times 3}$:
\State \quad $\tilde{F}[b, i, k] \gets$ feature for batch index $b$, qubit $i$, component $k$, for $i=0,\dots,n-1$, $k=0,1,2$
\Statex \textbf{(4) Rotation Mapping}
\For{$b = 0$ to $B-1$}
    \For{$i = 0$ to $n-1$}
        \State $\theta_{b,i}^{(x)} \gets \pi \cdot \tanh\big(\tilde{F}[b, i, 0]\big)$
        \State $\theta_{b,i}^{(y)} \gets \pi \cdot \tanh\big(\tilde{F}[b, i, 1]\big)$
        \State $\theta_{b,i}^{(z)} \gets \pi \cdot \tanh\big(\tilde{F}[b, i, 2]\big)$
    \EndFor
\EndFor
\State Pack angles into $\Theta \in \mathbb{R}^{B \times n \times 3}$ with entries
\State \quad $\Theta[b, i, :] = \big(\theta_{b,i}^{(x)}, \theta_{b,i}^{(y)}, \theta_{b,i}^{(z)}\big)$
\State \Return $\Theta$
\end{algorithmic}
\end{algorithm}

\subsection{Entanglement Patterns, Noise Modeling, and Measurement Layer}
We implement four hybrid models, each characterized by a distinct qubit entanglement pattern, building upon the design principles established by El Maouaki et al.~\citep{el2024advqunn}: (1) no entanglement, (2) linear entanglement, (3) star entanglement, and (4) full entanglement. 

In the subsequent sections, we detail the variational single-qubit rotation and entanglement blocks, followed by the formulation of the mixed noise block and the final measurement layer.

\subsubsection{No Entanglement Structure}
\begin{figure}
  \centering
    \includegraphics[width=0.7\textwidth]{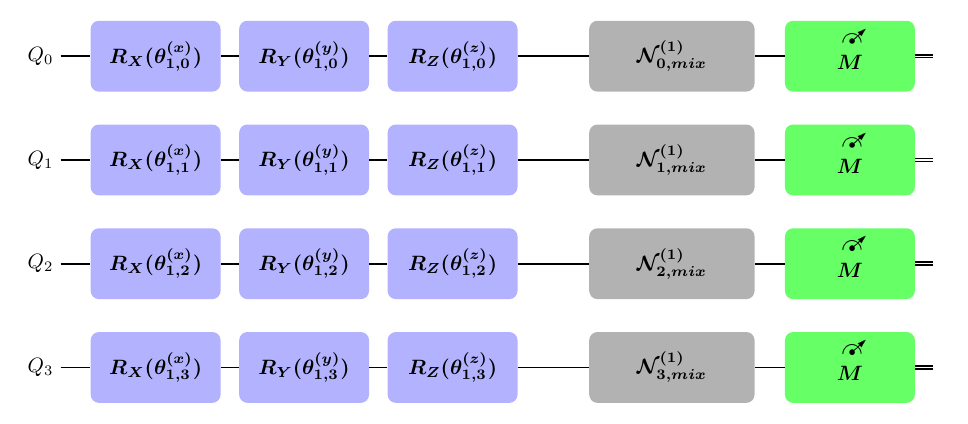}
    \caption{No-entanglement quantum circuit architecture. 
    Each qubit undergoes independent parameterized single-qubit rotations $R_X$, $R_Y$, and $R_Z$, subsequently followed by a local noise channel $\mathcal{N}^{(1)}_{i,\mathrm{mix}}$ and a final measurement $M$.}
    \label{fig:no_entanglement}
\end{figure}

As illustrated in Fig.~\ref{fig:no_entanglement}, the no-entanglement structure applies exclusively independent single-qubit operations, deliberately omitting all multi-qubit entangling gates. In this configuration, no controlled operations (e.g., CNOT gates) are executed; consequently, no quantum correlations are generated between qubits at any stage within the circuit. 

In our implementation, the circuit comprises a single variational layer (\(\ell = 1\)), which consists of a variational rotation block followed immediately by a mixed noise block, bypassing any intermediate entanglement stage. Although this model is constructed using qubits and quantum gate primitives, the complete absence of entanglement implies that it does not leverage this defining resource of quantum computation. Therefore, this configuration functions as a qubit-based, non-entangling baseline, included primarily to facilitate ablation studies and comparative analysis.

At the onset of the layer, independent parameterized single-qubit rotations are applied across all qubits. The rotation block is mathematically defined as
\begin{equation}
U_{\mathrm{rot}}(\boldsymbol{\vec{\theta}_{1}})
=
\bigotimes_{i=0}^{n-1}
\left(
R_{Z}\!\big(\theta^{(z)}_{1,i}\big)\,
R_{Y}\!\big(\theta^{(y)}_{1,i}\big)\,
R_{X}\!\big(\theta^{(x)}_{1,i}\big)
\right),
\end{equation}
where \(R_{\alpha}(\theta)=e^{-i\theta \sigma_{\alpha}/2}\) for \(\alpha\in\{X,Y,Z\}\), and \(\theta^{(x)}_{1,i}\), \(\theta^{(y)}_{1,i}\), and \(\theta^{(z)}_{1,i}\) denote the rotation parameters corresponding to qubit \(i\).

Given the absence of entangling operations, the ideal unitary transformation executed by the single variational layer reduces simply to
\begin{equation}
U_{1}(\boldsymbol{\vec{\theta}_{1}})
=
U_{\mathrm{rot}}(\boldsymbol{\vec{\theta}_{1}}).
\end{equation}

Following the rotation block, a mixed quantum noise channel $\mathcal{N}^{(1)}_{i,\mathrm{mix}}$ is applied independently to each qubit to accurately model hardware-induced imperfections. 

\subsubsection{Linear Entanglement Structure}
\begin{figure}
  \centering
    \includegraphics[width=0.85\textwidth]{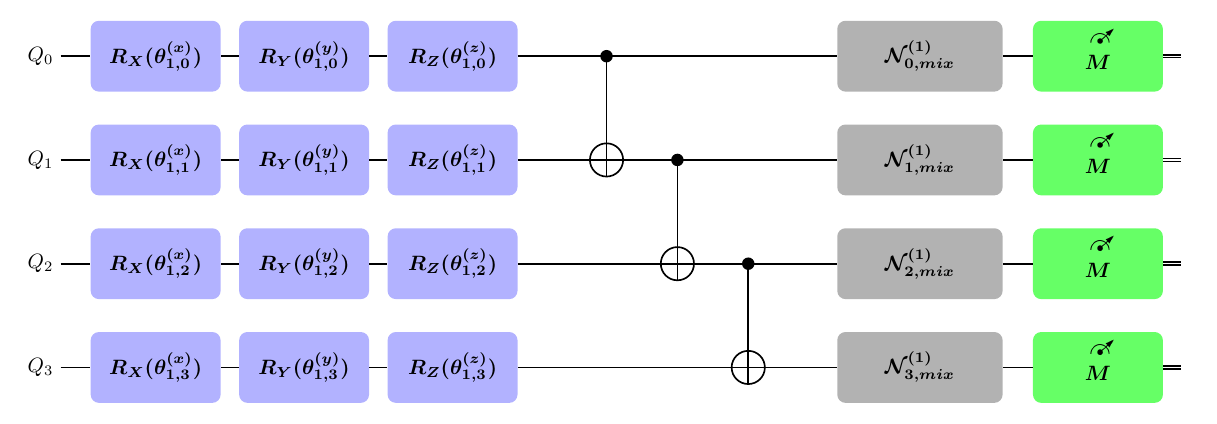}
    \caption{Linear entanglement quantum circuit architecture. 
    Each qubit is initially encoded via parameterized single-qubit rotations $R_X$, $R_Y$, and $R_Z$. 
    Neighboring qubits are subsequently entangled using controlled operations, establishing a chain-like connectivity. 
    Following entanglement, each qubit is subjected to a local noise channel $\mathcal{N}^{(1)}_{i,\mathrm{mix}}$ prior to the final measurement $M$.}
    \label{fig:linear_entanglement}
\end{figure}

As depicted in Fig.~\ref{fig:linear_entanglement}, the linear entanglement structure organizes the $n$ qubits into a one-dimensional chain, wherein entanglement is introduced exclusively between nearest neighbors. In our implementation, the circuit utilizes a single variational layer (\(\ell = 1\)) that adheres to a fixed processing pipeline: a variational rotation block, an entanglement block, and a subsequent mixed noise block.

At the onset of the layer, independent parameterized single-qubit rotations are applied to all qubits. The rotation block is mathematically defined as
\begin{equation}
U_{\mathrm{rot}}(\boldsymbol{\vec{\theta}_{1}})
=
\bigotimes_{i=0}^{n-1}
\left(
R_{Z}\!\big(\theta^{(z)}_{1,i}\big)\,
R_{Y}\!\big(\theta^{(y)}_{1,i}\big)\,
R_{X}\!\big(\theta^{(x)}_{1,i}\big)
\right),
\end{equation}
where \(R_{\alpha}(\theta)=e^{-i\theta \sigma_{\alpha}/2}\) for \(\alpha\in\{X,Y,Z\}\) denotes the single-qubit rotation, and \(\theta^{(x)}_{1,i}\), \(\theta^{(y)}_{1,i}\), and \(\theta^{(z)}_{1,i}\) represent the rotation angle parameters associated with qubit \(i\).

Following the rotation block, entanglement is generated by executing a sequence of CNOT gates between adjacent qubits along the chain. For a system of $n$ qubits, this entanglement block requires exactly $n-1$ two-qubit gates, establishing a linear connectivity pattern that scales efficiently with $O(n)$ entangling operations. The corresponding entangling unitary is given by
\begin{equation}
U_{\mathrm{linear\text{-}ent}}
=
\prod_{i=0}^{n-2} \mathrm{CNOT}_{i,i+1},
\end{equation}
where each $\mathrm{CNOT}_{i,i+1}$ operation entangles qubit $i$ with its neighbor, qubit $i+1$. This connectivity structure is treated as a fixed architectural hyperparameter and remains unchanged throughout the training process.

By combining the rotation and entanglement blocks, the ideal unitary transformation implemented by the single variational layer, prior to the application of the mixed noise block, is expressed as
\begin{equation}
\begin{aligned}
U_{1}(\boldsymbol{\vec{\theta}_{1}})
&=
U_{\mathrm{linear\text{-}ent}}\,
U_{\mathrm{rot}}(\boldsymbol{\vec{\theta}_{1}}) \\
&=
\left( \prod_{i=0}^{n-2} \mathrm{CNOT}_{i,i+1} \right)
\left(
\bigotimes_{i=0}^{n-1}
R_{Z}\!\big(\theta^{(z)}_{1,i}\big)\,
R_{Y}\!\big(\theta^{(y)}_{1,i}\big)\,
R_{X}\!\big(\theta^{(x)}_{1,i}\big)
\right).
\end{aligned}
\end{equation}

Following the entanglement block, a mixed quantum noise channel $\mathcal{N}^{(1)}_{i,\mathrm{mix}}$ is applied independently to each qubit to simulate realistic hardware-induced imperfections.

\subsubsection{Star Entanglement Structure}
\begin{figure}
  \centering
    \includegraphics[width=0.85\textwidth]{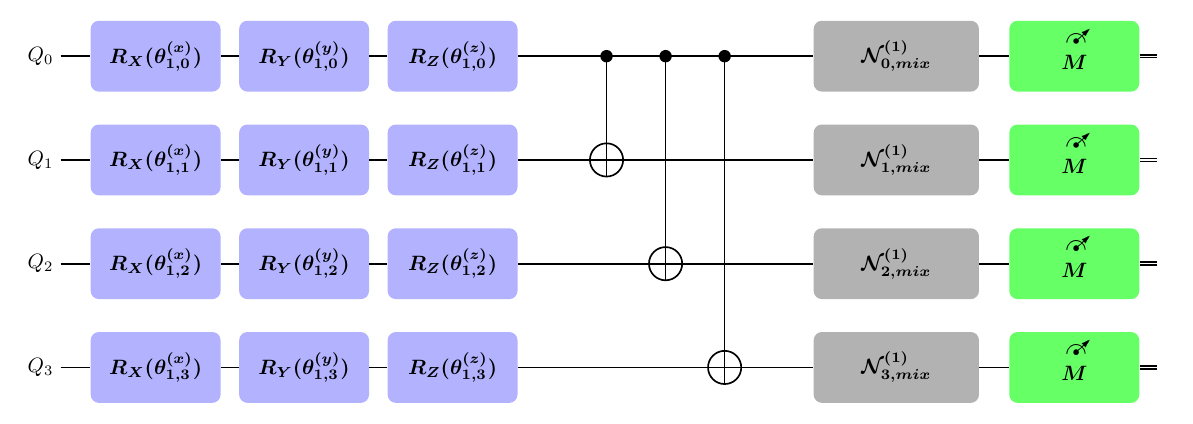}
    \caption{Star entanglement quantum circuit architecture. 
    Each qubit is initially encoded via parameterized single-qubit rotations $R_X$, $R_Y$, and $R_Z$. 
    Entanglement is then applied using a star topology, wherein a central hub qubit is connected to all peripheral qubits. Following entanglement, each qubit is subjected to a local noise channel $\mathcal{N}^{(1)}_{i,\mathrm{mix}}$ prior to the final measurement $M$.}
    \label{fig:star_entanglement}
\end{figure}

As depicted in Fig.~\ref{fig:star_entanglement}, the star entanglement structure organizes the $n$ qubits around a central hub qubit, designated as qubit $0$, which is subsequently entangled with all remaining peripheral qubits. In this topology, entanglement is generated by executing CNOT gates between the central hub and each peripheral node. For an $n$-qubit system, this configuration requires exactly $n-1$ two-qubit gates, yielding a hub-and-spoke connectivity pattern characterized by maximal fan-out centered on qubit $0$.

In our implementation, the circuit consists of a single variational layer (\(\ell = 1\)) that follows a fixed processing pipeline: a variational rotation block, a star-shaped entanglement block, and a subsequent mixed noise block.

At the beginning of the layer, independent parameterized single-qubit rotations are applied across all qubits. The rotation block is mathematically defined as
\begin{equation}
U_{\mathrm{rot}}(\boldsymbol{\vec{\theta}_{1}})
=
\bigotimes_{i=0}^{n-1}
\left(
R_{Z}\!\big(\theta^{(z)}_{1,i}\big)\,
R_{Y}\!\big(\theta^{(y)}_{1,i}\big)\,
R_{X}\!\big(\theta^{(x)}_{1,i}\big)
\right),
\end{equation}
where \(R_{\alpha}(\theta)=e^{-i\theta \sigma_{\alpha}/2}\) for \(\alpha\in\{X,Y,Z\}\), and \(\theta^{(x)}_{1,i}\), \(\theta^{(y)}_{1,i}\), and \(\theta^{(z)}_{1,i}\) represent the rotation parameters corresponding to qubit \(i\).

Following the rotation block, entanglement is introduced by coupling the central hub qubit to each peripheral qubit via CNOT gates. The corresponding entangling unitary is given by
\begin{equation}
U_{\mathrm{star\text{-}ent}}
=
\prod_{i=1}^{n-1} \mathrm{CNOT}_{0,i},
\end{equation}
where $\mathrm{CNOT}_{0,i}$ denotes a controlled-NOT gate with the control on qubit $0$ and the target on qubit $i$. This hub-and-spoke connectivity pattern is treated as a fixed architectural hyperparameter and remains unchanged throughout training.

By combining the rotation and entanglement blocks, the ideal unitary transformation implemented by the single variational layer, prior to the application of noise, is expressed as
\begin{equation}
\begin{aligned}
U_{1}(\boldsymbol{\vec{\theta}_{1}})
&=
U_{\mathrm{star\text{-}ent}}\,
U_{\mathrm{rot}}(\boldsymbol{\vec{\theta}_{1}}) \\
&=
\left( \prod_{i=1}^{n-1} \mathrm{CNOT}_{0,i} \right)
\left(
\bigotimes_{i=0}^{n-1}
R_{Z}\!\big(\theta^{(z)}_{1,i}\big)\,
R_{Y}\!\big(\theta^{(y)}_{1,i}\big)\,
R_{X}\!\big(\theta^{(x)}_{1,i}\big)
\right).
\end{aligned}
\end{equation}

After the entanglement block, a mixed quantum noise channel $\mathcal{N}^{(1)}_{i,\mathrm{mix}}$ is applied independently to each qubit to accurately model hardware-induced imperfections.

\subsubsection{Full Entanglement Structure}
\begin{figure}
  \centering
    \includegraphics[width=1.0\textwidth]{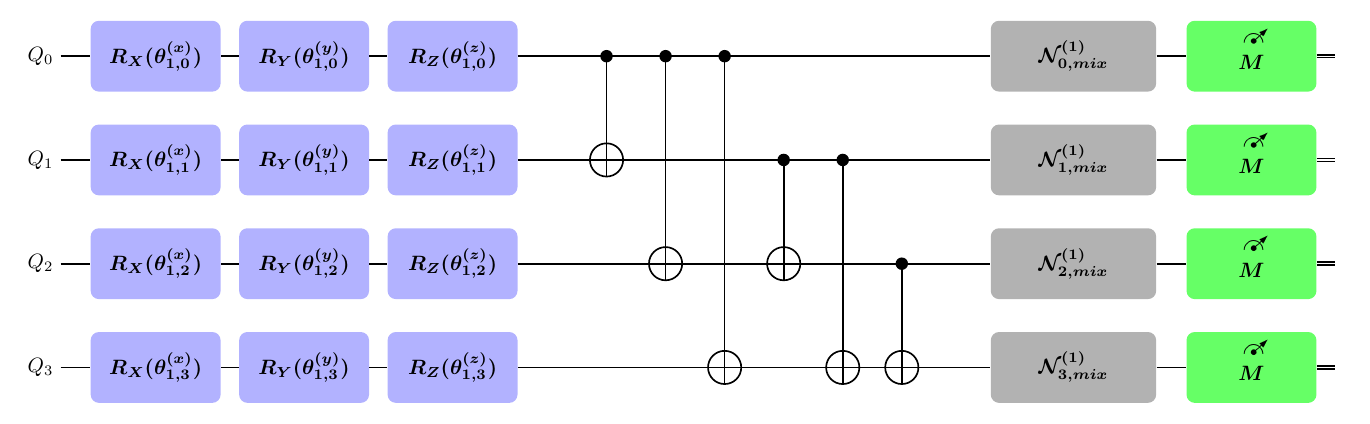}
    \caption{Full entanglement quantum circuit architecture. 
    Each qubit is initially encoded via parameterized single-qubit rotations $R_X$, $R_Y$, and $R_Z$. 
    Entanglement is subsequently applied between all possible pairs of qubits, establishing an all-to-all connectivity that maximizes shared correlations across the circuit. 
    Following entanglement, each qubit is subjected to a local noise channel $\mathcal{N}^{(1)}_{i,\mathrm{mix}}$ prior to the final measurement $M$.}
    \label{fig:full_entanglement}
\end{figure}

As illustrated in Fig.~\ref{fig:full_entanglement}, the full entanglement structure maximizes inter-qubit interactions by entangling every distinct pair of qubits within the system. For a register of $n$ qubits, this topology requires $\frac{n(n-1)}{2}$ two-qubit entangling operations, resulting in an all-to-all connectivity pattern. Although this structure provides maximal expressivity, it incurs a quadratic entangling cost, making it the most resource-intensive among the evaluated configurations.

In our implementation, the circuit consists of a single variational layer (\(\ell = 1\)) that follows a fixed processing pipeline: a variational rotation block, a full entanglement block, and a subsequent mixed noise block.

At the onset of the layer, independent parameterized single-qubit rotations are applied across all qubits. The rotation block is mathematically defined as
\begin{equation}
U_{\mathrm{rot}}(\boldsymbol{\vec{\theta}_{1}})
=
\bigotimes_{i=0}^{n-1}
\left(
R_{Z}\!\big(\theta^{(z)}_{1,i}\big)\,
R_{Y}\!\big(\theta^{(y)}_{1,i}\big)\,
R_{X}\!\big(\theta^{(x)}_{1,i}\big)
\right),
\end{equation}
where \(R_{\alpha}(\theta)=e^{-i\theta \sigma_{\alpha}/2}\) for \(\alpha\in\{X,Y,Z\}\), and \(\theta^{(x)}_{1,i}\), \(\theta^{(y)}_{1,i}\), and \(\theta^{(z)}_{1,i}\) represent the rotation parameters corresponding to qubit \(i\).

Following the rotation block, entanglement is introduced between all unordered pairs of qubits via CNOT gates. The corresponding entangling unitary is expressed as
\begin{equation}
U_{\mathrm{full\text{-}ent}}
=
\prod_{j=0}^{n-2}
\prod_{k=j+1}^{n-1}
\mathrm{CNOT}_{j,k},
\end{equation}
where $\mathrm{CNOT}_{j,k}$ denotes a controlled-NOT gate with the control on qubit $j$ and the target on qubit $k$. This all-to-all connectivity pattern is treated as a fixed architectural hyperparameter and remains unchanged throughout training.

By combining the rotation and entanglement blocks, the ideal unitary transformation executed by the single variational layer, prior to the application of noise, is given by
\begin{equation}
\begin{aligned}
U_{1}(\boldsymbol{\vec{\theta}_{1}})
&=
U_{\mathrm{full\text{-}ent}}\,
U_{\mathrm{rot}}(\boldsymbol{\vec{\theta}_{1}}) \\
&=
\left( \prod_{j=0}^{n-2} \prod_{k=j+1}^{n-1} \mathrm{CNOT}_{j,k} \right)
\left(
\bigotimes_{i=0}^{n-1}
R_{Z}\!\big(\theta^{(z)}_{1,i}\big)\,
R_{Y}\!\big(\theta^{(y)}_{1,i}\big)\,
R_{X}\!\big(\theta^{(x)}_{1,i}\big)
\right).
\end{aligned}
\end{equation}

After the full entanglement block, a mixed quantum noise channel $\mathcal{N}^{(1)}_{i,\mathrm{mix}}$ is applied independently to each qubit to simulate realistic hardware-induced imperfections. 

\subsubsection{Mixed Noise Channel}
Following each entanglement block, an independent single-qubit mixed noise channel is applied to every qubit to realistically simulate hardware imperfections. In this mixed setting, the local noise affecting qubit $i$ is modeled as a composition of three completely positive trace-preserving (CPTP) channels: depolarizing noise, amplitude damping, and phase damping. Depolarizing noise captures unbiased stochastic errors by randomly replacing the qubit state with the maximally mixed state, thereby modeling symmetric gate and control imperfections. Amplitude damping noise represents energy relaxation processes, such as spontaneous emission and dissipative scattering, which drive excited states toward the ground state. Phase damping noise models pure dephasing, describing the loss of quantum coherence without any corresponding exchange of energy \citep{sutojo2025acceptable,domingo2023taking}.

Let $\eta^{(\ell)} \in [0,1]$ denote the base noise strength associated with variational layer $\ell$, and let $w_D, w_A, w_P \ge 0$ be fixed mixing coefficients satisfying $w_D + w_A + w_P = 1$ (in our implementation, $w_D = 0.4$, $w_A = 0.3$, and $w_P = 0.3$). For a single-qubit density operator $\rho$, representing the state of qubit $i$ immediately following the ideal unitary at layer $\ell$, we define the mixed noise channel as
\begin{equation}
\mathcal{N}^{(\ell)}_{i,\mathrm{mix}}(\rho)
=
\left(
\mathcal{P}^{(w_P \eta^{(\ell)})}_i
\circ
\mathcal{A}^{(w_A \eta^{(\ell)})}_i
\circ
\mathcal{D}^{(w_D \eta^{(\ell)})}_i
\right)(\rho),
\end{equation}
where $\mathcal{D}^{(\cdot)}_i$, $\mathcal{A}^{(\cdot)}_i$, and $\mathcal{P}^{(\cdot)}_i$ denote the depolarizing, amplitude damping, and phase damping channels acting on qubit $i$, respectively. Each channel is CPTP, with effective noise parameters strictly constrained to the interval $[0,1]$. This formulation corresponds to the sequential application of all three noise processes immediately after the ideal unitary operation at each variational layer.

Under the assumption of independent local noise across qubits, the corresponding global mixed noise channel at layer $\ell$ factorizes as
\begin{equation}
\mathcal{N}^{(\ell)}_{\mathrm{global},\mathrm{mix}}
=
\bigotimes_{i=0}^{n-1}
\mathcal{N}^{(\ell)}_{i,\mathrm{mix}} \, .
\end{equation}

\subsubsection{Measurement Layer}
Following the variational rotation, entanglement, and mixed noise blocks, all $n$ qubits undergo projective measurement in the computational $Z$-basis. This process yields the full Born probability distribution:
\begin{equation}
\mathbf{p}^{(2^n)} = \big(\Pr(x)\big)_{x \in \{0,1\}^n} \in \Delta_{2^n-1} \subset \mathbb{R}^{2^n},
\end{equation}
where $x = (x_1,\dots,x_n)$ denotes the binary string labeling the computational basis state $\ket{x} = \ket{x_1 x_2 \cdots x_n}$, and $\Pr(x) = \bra{x}\rho\ket{x}$ represents the corresponding Born probability \citep{benedetti2019parameterized}.

Because the downstream classification task involves $K$ output classes, the $2^n$-dimensional probability vector is mapped to a class-level quantum probability vector $\mathbf{p}_q \in \Delta_{K-1} \subset \mathbb{R}^K$ via a fixed truncation-and-padding scheme. Specifically, we define an intermediate vector $\tilde{p}_q$:
\begin{equation}
\tilde{p}_q =
\begin{cases}
\big(\mathbf{p}^{(2^n)}_1, \ldots, \mathbf{p}^{(2^n)}_K\big), & \text{if } 2^n \ge K, \\[4pt]
\big(\mathbf{p}^{(2^n)}_1, \ldots, \mathbf{p}^{(2^n)}_{2^n}, \underbrace{0, \ldots, 0}_{K - 2^n}\big), & \text{if } 2^n < K.
\end{cases}
\end{equation}

To ensure numerical stability during normalization and to prevent division by zero, we introduce a small constant $\varepsilon = 10^{-8}$ in the denominator. Each quantum output component is normalized as
\begin{equation}
p_{q,k}
=
\frac{\tilde{p}_{q,k}}{\sum_{j=1}^{K} \tilde{p}_{q,j} + \varepsilon},
\qquad k = 1, \ldots, K .
\end{equation}
The final quantum probability vector is then constructed as
\begin{equation}
\mathbf{p}_q = \bigl[ p_{q,1}, \ldots, p_{q,K} \bigr].
\end{equation}

This procedure guarantees that ${p}_q$ lies strictly within the $(K-1)$-dimensional probability simplex, possessing non-negative entries that sum to one. These probabilities are subsequently converted to log-probabilities and fused with the classical CNN backbone output via the learnable dynamic weighting module to produce the final hybrid prediction.

\subsection{Dynamic Fusion Coefficient}
We propose a \emph{Dynamic Weighting Module} designed to adaptively fuse classical and quantum outputs within the hybrid architecture by exploiting statistical information derived from both modalities. This module is implemented as a lightweight multi-layer perceptron that predicts a continuous fusion coefficient $\alpha \in [0,1]$. This coefficient explicitly controls the relative contribution of the classical and quantum predictions in generating the final hybrid output.

Algorithm~\ref{alg:dynamic-fusion} formalizes the computation of the adaptive fusion coefficient $\alpha$. The method extracts high-order statistical features from the probability distributions produced by both the classical backbone and the quantum circuit, capturing model-specific confidence characteristics as well as cross-model agreement. By jointly modeling distributional sharpness, uncertainty, and mutual alignment, the Dynamic Weighting Module learns to adjudicate between the two predictive streams in a data-driven manner. This yields an adaptive fusion strategy that dynamically responds to the relative reliability of the classical and quantum components.

The process begins by calculating per-sample statistics for the classical and quantum probability distributions, denoted $p_c, p_q \in \Delta^{K-1} \subset \mathbb{R}^K$, which are produced by the classical backbone and the quantum circuit, respectively. For each modality, the algorithm computes the standard deviation $\sigma$ ($\sigma_c, \sigma_q$) to measure the spread of the distribution, the maximum probability $p^{\max}$ ($p_c^{\max}, p_q^{\max}$) to capture the peak prediction confidence, and the kurtosis $\kappa$ ($\kappa_c, \kappa_q$) to quantify the sharpness or dominance of the predicted class. These features provide a robust numerical representation of how certain each model is regarding its respective decision. To assess the consensus between the two models, the algorithm computes a cross-model agreement metric $\rho_{cq}$. Both distributions are mean-centered to produce $\tilde{p}_c$ and $\tilde{p}_q$, and their cosine similarity is calculated. This metric identifies whether the models are converging on a similar class ranking or providing conflicting signals, serving as a critical indicator for the fusion logic. A value approaching $1$ indicates strong agreement, $0$ implies independence, and negative values suggest disagreement between the two distributions.

These seven statistical features, three characterizing the classical backbone, three characterizing the quantum circuit, and one representing the cross-model agreement metric, are concatenated into a single feature vector $\mathbf{f} \in \mathbb{R}^7$. This unified representation jointly captures the internal predictive certainty of each model alongside their mutual alignment.

\begin{figure}
    \centering
    \includegraphics[width=0.99\textwidth, trim=15 25 15 20, clip]{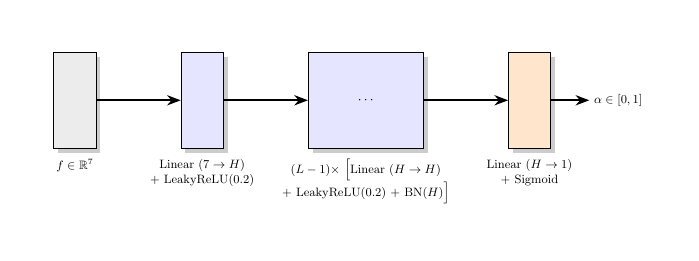}
    \caption{MLP architecture for fusion coefficient inference. The network generates the adaptive fusion coefficient $\alpha$ via a depth-$L$ MLP with a hidden width of $H$.}
    \label{fig:mlp-fusion}
\end{figure}

The feature vector $\mathbf{f}$ is subsequently processed by an MLP $g_\theta$ featuring a depth of $L=3$ and a hidden width of $H=128$. The network employs $\mathrm{LeakyReLU}$ activations to preserve gradient flow, in conjunction with $\mathrm{BatchNorm1d}$ layers to enhance numerical stability during training. The final layer applies a sigmoid activation, constraining the output to the interval $[0,1]$ and yielding the adaptive fusion coefficient $\alpha$. The architecture of the dynamic weighting MLP is illustrated in Fig.~\ref{fig:mlp-fusion}.

The fusion coefficient $\alpha$ dynamically adapts to the relative reliability of the models: it approaches $1$ when the classical predictions are highly confident and consistent with the quantum outputs, and it decreases when the quantum model provides a stronger, more reliable prediction.

\begin{algorithm}
\caption{Dynamic Weighting Module for Hybrid Fusion}
\label{alg:dynamic-fusion}
\begin{algorithmic}[1]
\Require Classical probability vector $p_c \in \Delta^{K-1}\subset\mathbb{R}^K$, quantum probability vector $p_q \in \Delta^{K-1}\subset\mathbb{R}^K$, stability constant $\epsilon>0$
\Require MLP $g_\theta:\mathbb{R}^7\rightarrow[0,1]$ with depth $L=3$, hidden width $H=128$ (LeakyReLU, BatchNorm1d, Sigmoid output)
\Ensure Adaptive fusion coefficient $\alpha \in [0,1]$
\Statex
\Statex \textbf{(1) Per-Modality Distribution Statistics}
\State $\mu_c \gets \frac{1}{K}\sum_{k=1}^{K} p_c^{(k)}$
\State $\sigma_c \gets \sqrt{\frac{1}{K}\sum_{k=1}^{K}\left(p_c^{(k)}-\mu_c\right)^2}$
\State $p_c^{\max} \gets \max_{k} p_c^{(k)}$
\State $\kappa_c \gets \dfrac{\frac{1}{K}\sum_{k=1}^{K}\left(p_c^{(k)}-\mu_c\right)^4}{\sigma_c^4+\epsilon}$
\State $\mu_q \gets \frac{1}{K}\sum_{k=1}^{K} p_q^{(k)}$
\algstore{dynamicfusion}
\end{algorithmic}
\end{algorithm}

\addtocounter{algorithm}{-1}

\begin{algorithm}
\caption{(continued)}
\begin{algorithmic}[1]
\algrestore{dynamicfusion}
\State $\sigma_q \gets \sqrt{\frac{1}{K}\sum_{k=1}^{K}\left(p_q^{(k)}-\mu_q\right)^2}$
\State $p_q^{\max} \gets \max_{k} p_q^{(k)}$
\State $\kappa_q \gets \dfrac{\frac{1}{K}\sum_{k=1}^{K}\left(p_q^{(k)}-\mu_q\right)^4}{\sigma_q^4+\epsilon}$
\Statex \textbf{(2) Cross-Model Agreement}
\State $\tilde{p}_c \gets p_c - \mu_c\mathbf{1}$
\State $\tilde{p}_q \gets p_q - \mu_q\mathbf{1}$
\State $\rho_{cq} \gets \dfrac{\tilde{p}_c^\top \tilde{p}_q}{\|\tilde{p}_c\|_2\,\|\tilde{p}_q\|_2+\epsilon}$ \Comment{$\rho_{cq}\in[-1,1]$}
\Statex \textbf{(3) Feature Vector Construction}
\State $\mathbf{f} \gets \big[\sigma_c,\; p_c^{\max},\; \kappa_c,\; \sigma_q,\; p_q^{\max},\; \kappa_q,\; \rho_{cq}\big] \in \mathbb{R}^7$
\Statex \textbf{(4) Adaptive Weight Prediction}
\State $\alpha \gets g_\theta(\mathbf{f})$ \Comment{Sigmoid output ensures $\alpha\in[0,1]$}
\State \Return $\alpha$
\Statex \vspace{0.25em}
\Statex \textbf{MLP Specification $g_\theta$ (used in Step 4):}
\Statex \hspace{0.75em}Input: $\mathrm{Linear}(7\rightarrow H)$ $\rightarrow$ $\mathrm{LeakyReLU}(0.2)$
\Statex \hspace{0.75em}Hidden (repeat $L-1$ times): $\mathrm{Linear}(H\rightarrow H)$ $\rightarrow$ $\mathrm{LeakyReLU}(0.2)$ $\rightarrow$ $\mathrm{BatchNorm1d}(H)$
\Statex \hspace{0.75em}Output: $\mathrm{Linear}(H\rightarrow 1)$ $\rightarrow$ $\mathrm{Sigmoid}$
\end{algorithmic}
\end{algorithm}

\subsection{Hybrid Output Fusion}
The final hybrid prediction is obtained by combining the classical and quantum predictive streams using the adaptive fusion coefficient $\alpha$ inferred by the Dynamic Weighting Module. Algorithm~\ref{alg:hybrid_fusion} formalizes the complete fusion pipeline, encompassing probability normalization, amplitude-space mixing, and the reconstruction of hybrid logits to facilitate standard negative log-likelihood loss optimization.

The fusion process begins by transforming the outputs of the classical and quantum components into directly comparable probability distributions over the $K$ classes. The classical backbone produces logits $\mathbf{y}_c$, which are converted into probabilities via the softmax operation. Concurrently, the quantum circuit produces an intermediate real-valued vector $\tilde{\mathbf{p}}_q$, which is first projected onto the nonnegative orthant to enforce physical validity. This vector is then normalized by its total mass, incorporating an additive stability constant $\epsilon$, to yield a valid probability distribution. Finally, both the classical and quantum probability vectors are element-wise clamped by $\epsilon$ to prevent numerical instabilities during subsequent logarithmic and square-root operations.

Next, following the methodology proposed by Cuéllar et al.~\citep{cuellar2023time}, both probability vectors are mapped into an amplitude representation via element-wise square roots. This transformation facilitates a smooth interpolation mechanism, wherein contributions combine linearly at the amplitude level prior to squaring, conceptually analogous to the mixing of wavefunction magnitudes.

A key design consideration in the fusion stage is the selection of the quantum scaling coefficient. Rather than employing a linear complementary relation such as $\beta = 1 - \alpha$, which fails to preserve normalization after squaring and can artificially amplify or suppress specific classes, we adopt an orthogonal scaling strategy. Specifically, the learned fusion coefficient $\alpha \in [0,1]$ dictates the classical contribution, while the quantum contribution is assigned a complementary coefficient $\beta$ defined as
\begin{equation}
\beta = \sqrt{\max(\epsilon, 1 - \alpha^2)} ,
\end{equation}
where $\epsilon$ is a small constant introduced to ensure numerical stability.

This construction strictly enforces $\alpha^2 + \beta^2 \approx 1$, yielding a norm-preserving mixture at the amplitude level. The resulting cosine--sine parameterization mirrors the normalization structure inherent to quantum state amplitudes, where superpositions satisfy the condition $|\alpha|^2 + |\beta|^2 = 1$. Although the classical and quantum components are not quantum states themselves, this geometric consistency ensures that their combined amplitudes remain balanced and well-conditioned.

The orthogonal fusion scheme offers several practical advantages. It enables a smooth and symmetric transition between classical and quantum regimes, prevents premature collapse to a single information source, and guarantees that neither component vanishes abruptly. For small values of $\alpha$, the quantum contribution dominates while the classical signal remains active; conversely, for large values of $\alpha$, the fusion is driven primarily by the classical backbone, with the quantum contribution gradually diminishing.

Finally, the fused amplitudes are squared to reconstruct a valid hybrid probability distribution, which is then renormalized. The hybrid logits are subsequently obtained by applying a logarithm to the reconstructed distribution, yielding a representation fully compatible with the negative log-likelihood loss utilized during training. Crucially, maintaining orthogonality prior to squaring is essential to prevent distortion in the final probabilities and to ensure stable, well-behaved hybrid fusion.

\begin{algorithm}
\caption{Hybrid Output Fusion}
\label{alg:hybrid_fusion}
\begin{algorithmic}[1]
\Require Classical logits $\mathbf{y}_c \in \mathbb{R}^K$, intermediate quantum vector $\tilde{\mathbf{p}}_q \in \mathbb{R}^K$,
        fusion coefficient $\alpha \in [0,1]$, stability constant $\epsilon>0$
\Ensure Hybrid logits $\mathbf{y}_{\mathrm{hybrid}} \in \mathbb{R}^K$
\Statex
\Statex \textbf{(1) Probability Normalization}
\State $\mathbf{p}_c \gets \mathrm{softmax}(\mathbf{y}_c)$
\State $\tilde{\mathbf{p}}_q \gets \max(\tilde{\mathbf{p}}_q,\mathbf{0})$ \Comment{Ensure nonnegativity}
\State $Z_q \gets \sum_{j=1}^{K} \tilde{p}_{q,j} + \epsilon$
\For{$k = 1$ \textbf{to} $K$}
    \State $p_{q,k} \gets \tilde{p}_{q,k} / Z_q$
\EndFor
\algstore{hybrid_fusion}
\end{algorithmic}
\end{algorithm}

\addtocounter{algorithm}{-1}

\begin{algorithm}
\caption{(continued)}
\begin{algorithmic}[1]
\algrestore{hybrid_fusion}
\State $\mathbf{p}_q \gets [p_{q,1},\ldots,p_{q,K}]^\top$
\State $\mathbf{p}_c \gets \max(\mathbf{p}_c,\epsilon)$;\;\; $\mathbf{p}_q \gets \max(\mathbf{p}_q,\epsilon)$ \Comment{Element-wise clamp}
\Statex \textbf{(2) Amplitude Representation}
\State $\mathbf{a}_c \gets \sqrt{\mathbf{p}_c}$;\;\; $\mathbf{a}_q \gets \sqrt{\mathbf{p}_q}$ \Comment{Element-wise}
\Statex \textbf{(3) Complementary Fusion Weight}
\State $\beta \gets \sqrt{\max(\epsilon,\,1-\alpha^2)}$ \Comment{So $\alpha^2+\beta^2\approx 1$}
\Statex \textbf{(4) Amplitude Fusion}
\State $\mathbf{a}_{\mathrm{hybrid}} \gets \alpha\,\mathbf{a}_c + \beta\,\mathbf{a}_q$
\Statex \textbf{(5) Probability Reconstruction}
\State $\hat{\mathbf{p}}_{\mathrm{hybrid}} \gets \mathbf{a}_{\mathrm{hybrid}} \odot \mathbf{a}_{\mathrm{hybrid}}$ \Comment{Element-wise square}
\State $Z_h \gets \sum_{j=1}^{K} \hat{p}_{\mathrm{hybrid},j} + \epsilon$
\For{$k = 1$ \textbf{to} $K$}
    \State $p_{\mathrm{hybrid},k} \gets \hat{p}_{\mathrm{hybrid},k} / Z_h$
\EndFor
\State $\mathbf{p}_{\mathrm{hybrid}} \gets [p_{\mathrm{hybrid},1},\ldots,p_{\mathrm{hybrid},K}]^\top$
\State $\mathbf{p}_{\mathrm{hybrid}} \gets \max(\mathbf{p}_{\mathrm{hybrid}},\epsilon)$ \Comment{Element-wise clamp}
\Statex \textbf{(6) Logit Recovery}
\State $\mathbf{y}_{\mathrm{hybrid}} \gets \log(\mathbf{p}_{\mathrm{hybrid}})$ \Comment{Element-wise log}
\State \Return $\mathbf{y}_{\mathrm{hybrid}}$
\end{algorithmic}
\end{algorithm}


\section{Experimental Setup}
\subsection{Datasets and Software Frameworks}
In this work, we evaluate the performance and adversarial robustness of classical neural networks alongside our proposed hybrid quantum--classical neural network architecture using three benchmark datasets: MNIST \citep{deng2012mnist,torchvision_mnist}, OrganAMNIST \citep{yang2023medmnist}, and CIFAR-10 \citep{lv2020cifar,torchvision_cifar10}. Furthermore, we utilize dedicated software frameworks to implement the quantum circuits and conduct the adversarial robustness evaluations. The quantum circuits underlying QShield are constructed and integrated via the PennyLane library \citep{bergholm2018pennylane}, which facilitates seamless hybrid quantum--classical modeling. To rigorously assess adversarial robustness, we employ both Torchattacks \citep{kim2020torchattacks} and the Adversarial Robustness Toolbox (ART) \citep{nicolae2018adversarial}, which offer comprehensive implementations of various adversarial attacks. Together, these datasets and frameworks establish the foundation of our experimental setup, enabling an extensive evaluation of model resilience under diverse adversarial conditions.

\subsubsection{Datasets}
We use three widely adopted benchmark datasets that span different visual domains: handwritten digits (MNIST), medical imaging (OrganAMNIST), and natural images (CIFAR-10). This selection enables evaluation across varying levels of visual complexity.

\begin{figure}
    \centering
    \includegraphics[width=0.86\textwidth]{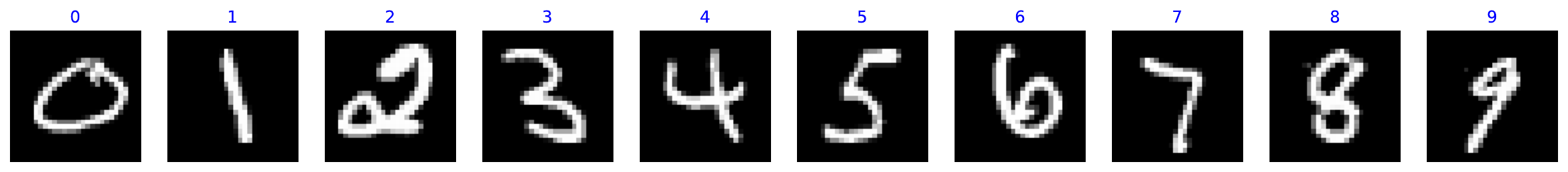}
    \caption{Sample images from the MNIST dataset. 
    The dataset consists of grayscale handwritten digits ranging from 0 to 9, each represented as a $28 \times 28$ pixel image. 
    Shown here is one example per class, with the corresponding ground-truth label displayed above each digit.}
    \label{fig:mnist_samples}
\end{figure}

\paragraph{MNIST}
This dataset consists of 60,000 training and 10,000 test grayscale images of handwritten digits, each sized at $28 \times 28$ pixels. It serves as a standard benchmark for image classification tasks. Representative samples are displayed in Fig.~\ref{fig:mnist_samples}.

\begin{figure}
    \centering
    \includegraphics[width=0.95\textwidth]{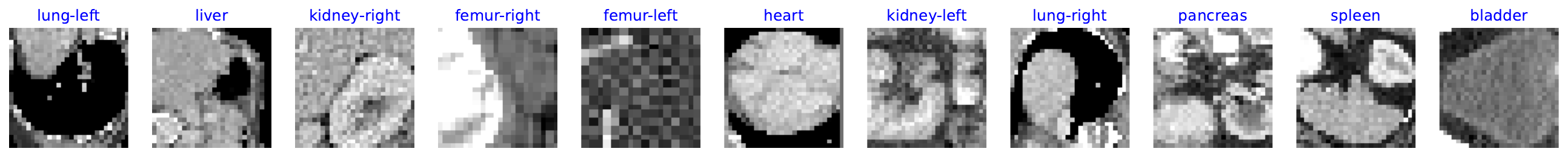}
    \caption{Sample images from the OrganAMNIST dataset. 
    The dataset contains grayscale abdominal CT slices annotated with organ labels. 
    Shown here is one representative $28 \times 28$ pixel image from each of the 11 classes (e.g., spleen, liver, kidneys, lungs, heart, pancreas, bladder, and femurs), with the ground-truth label displayed above each sample.}
    \label{fig:organamnist_samples}
\end{figure}

\paragraph{OrganAMNIST}
This dataset is a subset of the MedMNIST collection, containing 58,830 grayscale CT slices of 11 abdominal organs. The images are resized to $28 \times 28$ pixels and partitioned into training, validation, and test sets. It is commonly used for medical image classification tasks. Representative samples are presented in Fig.~\ref{fig:organamnist_samples}.

\begin{figure}
    \centering
    \includegraphics[width=0.94\textwidth]{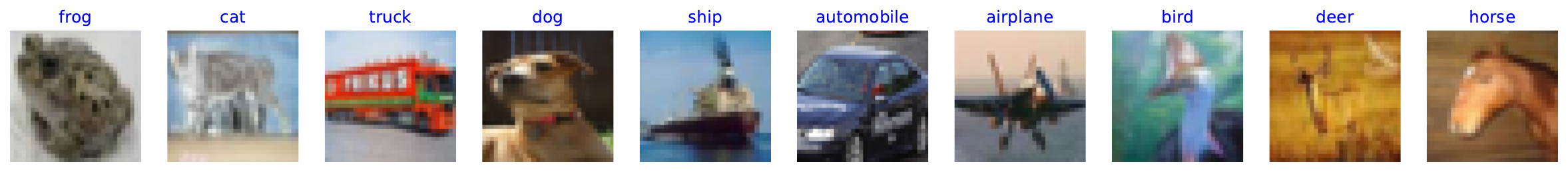}
    \caption{Sample images from the CIFAR-10 dataset. 
    The dataset consists of $32 \times 32$ color images across 10 object categories, including animals (e.g., cat, dog, horse, bird, deer, frog) and vehicles (e.g., airplane, automobile, ship, truck). 
    Shown here is one representative image per class, with the ground-truth label displayed above each sample.}
    \label{fig:cifar10_samples}
\end{figure}

\paragraph{CIFAR-10}
This dataset consists of 60,000 color images, each $32 \times 32$ pixels in size, categorized into 10 object classes, with 50,000 training and 10,000 test samples. It is a standard benchmark for natural image classification. Representative samples are illustrated in Fig.~\ref{fig:cifar10_samples}.

\subsubsection{Software Frameworks}
Our implementation relies on PennyLane for constructing and executing the hybrid quantum--classical models, while the adversarial robustness evaluation is conducted using Torchattacks and the Adversarial Robustness Toolbox. These libraries provide standardized implementations of common adversarial attack methods. Further details regarding each framework are provided below.

\paragraph{PennyLane}
PennyLane is a versatile, open-source Python library designed for quantum computing, quantum machine learning, and quantum chemistry. It enables users to construct and execute quantum circuits on both simulators and physical quantum hardware. Its seamless integration with popular machine learning frameworks, such as PyTorch and NumPy, facilitates the development of hybrid quantum--classical models for advanced applications.

\paragraph{Torchattacks}
Torchattacks is an open-source Python library dedicated to generating adversarial attacks on deep learning models. It is widely used to assess and improve model robustness. By supporting a variety of attack methods, it enables comprehensive testing of a model's resilience against adversarial inputs. Torchattacks offers extensive customization of attack parameters and integrates directly with PyTorch, providing an intuitive API for the implementation and evaluation of adversarial attacks.

\paragraph{Adversarial Robustness Toolbox}
The Adversarial Robustness Toolbox (ART) is a comprehensive Python library for evaluating and defending machine learning models against adversarial threats, including evasion, poisoning, extraction, and inference attacks. ART supports a broad spectrum of attacks and is compatible with major frameworks such as PyTorch and TensorFlow.

\subsection{Adversarial Threat Models and Attack Specifications}
To thoroughly evaluate model robustness, we consider both white-box and black-box adversarial threat models. These complementary settings facilitate assessment under direct access and query-limited attack scenarios. In our experiments, we employ representative attacks from each category, ranging from fast single-step perturbations to iterative and query-based methods. This experimental design ensures that robustness is evaluated across a broad spectrum of realistic adversarial conditions rather than being limited to a single threat model.

\subsubsection{White-box Adversarial Attacks}
White-box adversarial attacks operate under the assumption that the adversary possesses full access to the target model, including its architecture, parameters, and gradients \citep{li2024survey}. This strong threat model allows the attacker to directly compute the gradients of the loss function with respect to the input and construct adversarial perturbations that maximize prediction error \citep{feng2024comparative}. In this work, we employ several representative white-box attacks, which are detailed below.

\paragraph{Fast Gradient Sign Method (FGSM)} 
FGSM is a one-step gradient-based attack that perturbs the input in the direction of the sign of the loss gradient. By adding a small vector of magnitude $\epsilon$, whose elements correspond to the sign of the gradient with respect to the input, FGSM effectively drives the input toward a region that is more detrimental to the model's performance \citep{goodfellow2014explaining}. This straightforward method can induce misclassifications via imperceptible perturbations, causing the model to make incorrect predictions with high confidence.

\paragraph{Projected Gradient Descent (PGD)} 
PGD is an iterative, multi-step extension of FGSM that takes multiple small gradient steps, projecting the perturbed example back onto the allowed norm ball (e.g., an $L_\infty$ or $L_2$ budget) after each step to ensure the perturbation remains within specified bounds. PGD is widely regarded as the strongest attack that uses first-order (gradient) information, reliably discovering adversarial examples within the defined perturbation limit. As argued by Madry et al. \citep{madry2017towards}, PGD can be viewed as a universal adversary for a given norm, frequently serving as the baseline for evaluating adversarial robustness.

\paragraph{Auto-PGD (APGD)}  
APGD is an advanced variant of PGD that automatically adapts its step sizes and objectives, eliminating the need for manual hyperparameter tuning. Croce and Hein \citep{croce2020reliable} observed that employing a fixed step size and the standard cross-entropy loss can cause PGD to underestimate a model's vulnerability. APGD mitigates these issues by auto-tuning the step size at each iteration and utilizing an alternative loss function. This results in a parameter-free attack (excluding the number of iterations) that provides a more reliable evaluation of robustness. Consequently, APGD automates the configuration of PGD to consistently generate strong adversarial examples without requiring user-specified step sizes.

\paragraph{Variance Momentum Iterative FGSM (VMI-FGSM)} 
VMI-FGSM is a momentum-based iterative attack enhanced with variance tuning to improve its efficacy against defended or unknown models. Proposed by Wang \& He \citep{wang2021enhancing}, this method aims to boost transferability: at each iteration, rather than merely accumulating the current gradient in the momentum term, VMI-FGSM incorporates the gradient variance from previous iterations to adjust the update. This variance tuning stabilizes the update direction and helps the attack in escaping suboptimal local minima. Empirical results on ImageNet demonstrate that integrating gradient variance in this manner significantly enhances the success rate of attacks when transferred to other models, a common scenario in black-box settings.

\paragraph{Carlini \& Wagner (C\&W) Attack} 
The C\&W attack encompasses a family of optimization-based attacks introduced by Carlini and Wagner \citep{carlini2017towards} that seek the smallest perturbation required to induce misclassification. The most prominent variant is the $L_2$ C\&W attack, which formulates the search for an adversarial example as a constrained optimization problem, minimizing the perturbation norm while achieving a target misclassification, and solves it via iterative optimization. The C\&W $L_2$ attack is considered one of the most effective white-box attacks for the $L_2$ norm and is highly recommended as a primary method for evaluating defenses \citep{guesmi2022room}.

\paragraph{DeepFool Attack} 
DeepFool is an attack that iteratively linearizes the classifier to determine the minimal perturbation necessary to alter the decision. Introduced by Moosavi-Dezfooli et al. \citep{moosavi2016deepfool}, DeepFool initializes at a given input and advances toward the closest class boundary by assuming the model behaves approximately linearly within a small neighborhood. It generates perturbations that are often orders of magnitude smaller than those produced by FGSM for the same image. In practice, DeepFool discovers perturbations that are hardly perceptible, whereas FGSM typically outputs a perturbed image with a higher norm. Consequently, DeepFool serves as a valuable tool for measuring a classifier's minimum vulnerability to adversarial examples in terms of perturbation magnitude.

\subsubsection{Black-box Adversarial Attacks} 
Black-box adversarial attacks operate under the assumption that the adversary lacks access to the internal structure, parameters, or gradients of the target model \citep{li2024survey}. Instead, the attacker can only interact with the model via its outputs, such as predicted labels or confidence scores, or exploit the transferability of adversarial examples across distinct models \citep{mahmood2021back}. In this work, we employ two representative black-box attacks, which are described below.

\paragraph{One-Pixel Attack} 
The One-Pixel Attack is a highly constrained adversarial attack that modifies only a single pixel (or a very small number of pixels) in the input image. Su et al. \citep{su2019one} demonstrated that altering even a single pixel's color can successfully deceive deep neural networks on specific images. Since gradient information is unavailable in a pure black-box setting, the one-pixel attack utilizes an evolutionary strategy to identify the optimal pixel location and color modification that maximizes the model's prediction error or changes its label. Formally, the attack optimizes the target class probability subject to an $L_0$ constraint on the perturbation (with $d=1$ pixel changed). Despite its simplicity, this attack highlights that neural networks can occasionally be compromised by an almost imperceptible, single-pixel alteration.

\paragraph{Square Attack} 
The Square Attack is a query-efficient, score-based black-box attack that introduces random localized perturbations in the form of squares. Developed by Andriushchenko et al. \citep{andriushchenko2020square}, the Square Attack operates without relying on any gradient information, rendering it immune to defenses that mask gradients, and instead employs a randomized search strategy. At each iteration, it places a square patch of noise at a random location on the image and adjusts its intensity such that the overall perturbation remains precisely at the allowed norm threshold. By strategically selecting these square-shaped updates, the attack efficiently identifies an adversarial example using significantly fewer queries than naive random sampling. The Square Attack has demonstrated high success rates with low query counts, occasionally outperforming certain white-box attacks in specific contexts. This establishes it as a formidable representative of black-box attacks, particularly under strict query limitations.

\subsection{Attack Specifications}
Below, we outline the rationale underlying the selection of specific hyperparameters for the adversarial attacks. Comprehensive specifications of the chosen hyperparameters for each dataset and attack are provided in Appendix~\ref{appendix:attack_params}.

\subsubsection{Perturbation Budgets ($\varepsilon$)}
The $\epsilon$ values are dataset-specific and are selected to balance attack effectiveness with perceptual imperceptibility. MNIST tolerates larger perturbations due to its binary nature, whereas CIFAR-10 requires smaller perturbations to preserve visual fidelity.

\subsubsection{Optimization Parameters} 
Step sizes and iteration counts are calibrated based on convergence analysis and computational efficiency. In iterative adversarial attacks, random initialization is employed to introduce variability into the perturbation process, which enhances attack diversity and yields higher overall success rates.

\subsubsection{Reproducibility}
All experiments utilize fixed random seeds where applicable. Parameter configurations align with the standard implementations provided in the Torchattacks and Adversarial Robustness Toolbox (ART) libraries.

\subsubsection{Attack-Specific Adjustments}  
Certain hyperparameters are fine-tuned to accommodate the unique mechanics of each attack:
\begin{itemize}
    \item \textbf{C\&W Attack.} Binary search steps are scaled according to dataset complexity to balance optimization accuracy with runtime.
    \item \textbf{Square Attack.} Query budgets are selected to maximize attack strength while adhering to computational constraints.
    \item \textbf{One-Pixel Attack.} The population size is tuned to stabilize convergence within the discrete optimization process.
\end{itemize}

\subsection{Training and Validating the Models}
All experiments were conducted on a machine equipped with an NVIDIA GeForce MX570A GPU featuring 2 GB of memory. The entire hybrid architecture, including the classical backbone, quantum module, and adaptive fusion components, was trained end-to-end using a unified optimization objective. Models were trained for 10 epochs with a batch size of 128. Optimization was performed using the Adam optimizer applied jointly to all trainable parameters, and the training objective was defined by the Negative Log-Likelihood (NLL) loss computed on the final hybrid outputs.

\subsubsection{Training and Validation}
\begin{figure}
  \centering
  \includegraphics[width=0.95\textwidth]{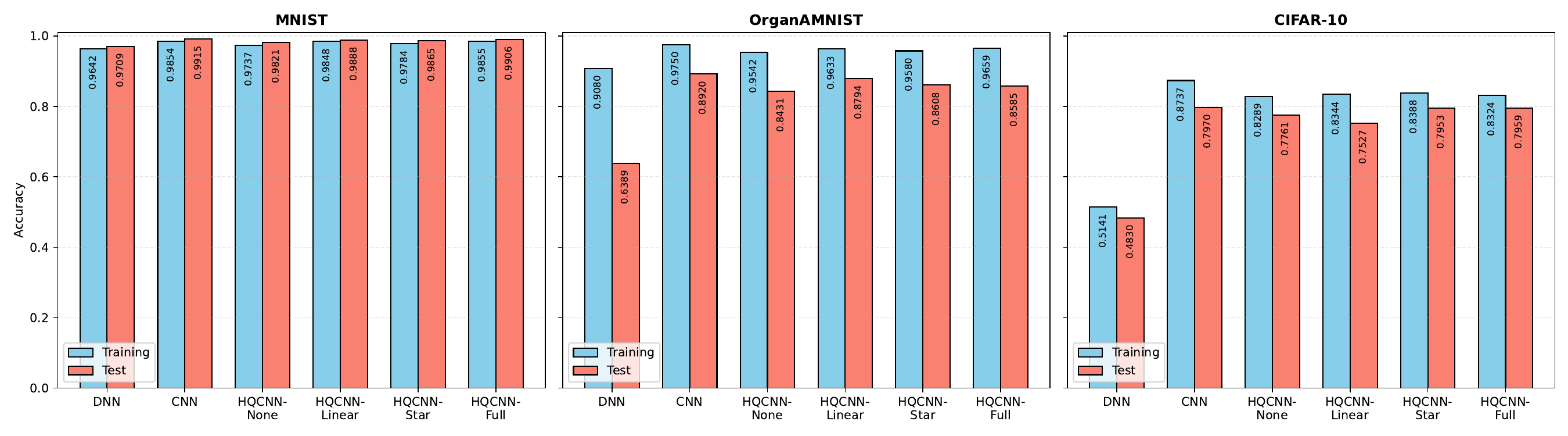}
    \caption{Training and test accuracies across datasets. 
    Comparison of classical (DNN, CNN) and quantum-enhanced (HQCNNs with no, linear, star, and full entanglement) models on MNIST, OrganAMNIST, and CIFAR-10. 
    While CNN achieves the highest overall accuracy, HQCNN variants exhibit comparable performance.}
  \label{fig:tva_results}
\end{figure}

Figure~\ref{fig:tva_results} presents the training and validation accuracies across all datasets and model variants. The CNN baseline consistently achieves the highest validation accuracy, while the HQCNN variants perform competitively, trailing the CNN by only marginal differences. Conversely, the DNN baseline exhibits substantially weaker performance and clear signs of overfitting, particularly on the OrganAMNIST dataset.

Both the CNN and HQCNN models demonstrate strong generalization, exhibiting modest train--test gaps across MNIST, OrganAMNIST, and CIFAR-10. This indicates that the integration of quantum components does not adversely impact stability or generalization. Overall, the CNN remains the strongest baseline, while the HQCNNs achieve near-parity performance and maintain stable training behavior.

\subsubsection{Total Time Cost}
\begin{figure}
  \centering
  \includegraphics[width=0.95\textwidth]{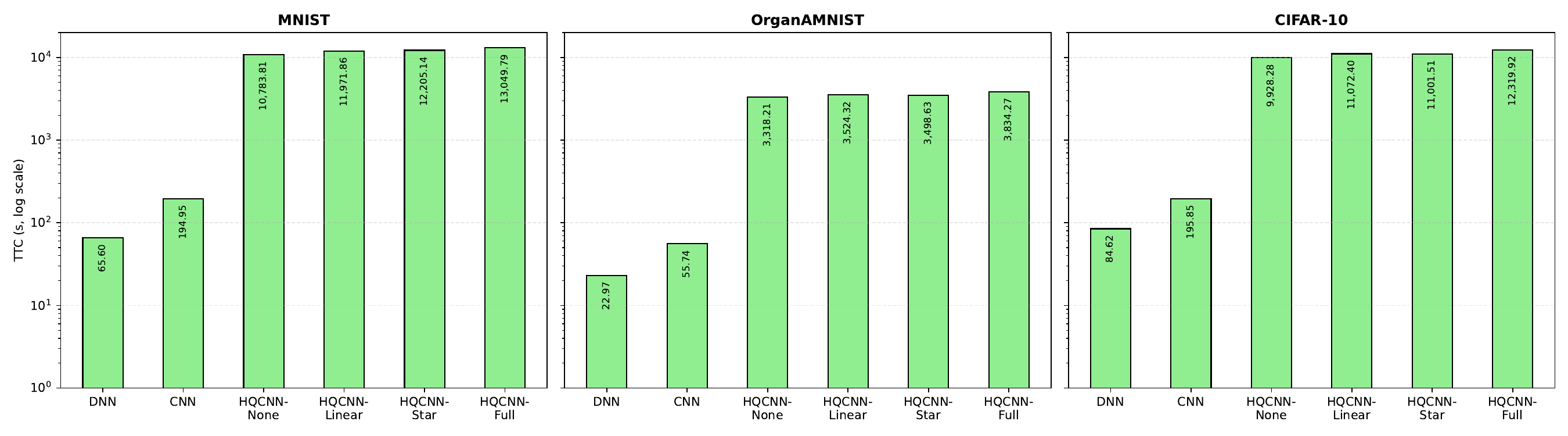}
    \caption{Total Time Cost (TTC) across datasets. 
    Measured computational cost (log-scaled seconds) for DNN, CNN, and HQCNN variants on MNIST, OrganAMNIST, and CIFAR-10. 
    Classical models (DNN, CNN) achieve substantially lower training times, while HQCNNs incur higher costs due to circuit simulation and evaluations plus entanglement operations, with complexity increasing from no to full entanglement.}
  \label{fig:ttc_results}
\end{figure}

Figure~\ref{fig:ttc_results} presents the total time cost for all model variants on a logarithmic scale. Classical models exhibit substantially lower computational costs, with CNNs being only moderately slower than DNNs. In contrast, HQCNN variants incur a significantly higher time cost, approximately two orders of magnitude greater, despite achieving comparable predictive accuracy.

It is crucial to note, however, that total time cost is an inherently imperfect metric for directly comparing classical and quantum-enhanced models. The reported HQCNN timings reflect the simulation of quantum circuits on classical hardware, where repeated circuit evaluations and entanglement operations heavily dominate the runtime. These costs do not directly correspond to execution times on future fault-tolerant quantum hardware, nor do they capture fundamental differences in operational expense, parallelism, or energy consumption between classical and quantum platforms.

Within this constrained but practical evaluation setting, CNNs remain the most efficient choice for time-critical applications. HQCNNs, by contrast, are more appropriately viewed as models that trade computational efficiency for potential quantum-specific advantages, such as improved adversarial robustness, rather than serving as immediate, drop-in replacements for classical architectures.

\section{Results}
\subsection{Adversarial Robustness Benchmarking}
We evaluate the adversarial robustness of six models: a Deep Neural Network (DNN), a Convolutional Neural Network (CNN), and four hybrid quantum--classical neural networks (HQCNNs) featuring distinct entanglement patterns. These evaluations are conducted across three benchmark datasets: MNIST, OrganAMNIST, and CIFAR-10. These datasets span varying levels of visual and structural complexity, enabling a comprehensive assessment of robustness under diverse conditions.

Each model is subjected to eight adversarial attacks: FGSM, PGD, APGD, VMI-FGSM, C\&W, DeepFool, One-Pixel, and Square Attack. Robustness is quantified using the Attack Success Rate (ASR), defined as the fraction of originally correct predictions that are successfully altered by adversarial perturbations. For reference, the Original Detection Rate (ODR), which corresponds to the clean test accuracy prior to any attack, is reported alongside each model in the accompanying figures.

By jointly analyzing the ODR and ASR across datasets and attack methodologies, we disentangle the trade-off between standard predictive performance and adversarial robustness. This unified benchmarking framework enables a systematic comparison between classical and quantum-enhanced architectures, highlighting their respective strengths and limitations under established adversarial threat models.

\subsubsection{MNIST Dataset}
\paragraph{Model-specific observations}
On clean MNIST data, the DNN attains an ODR of $97.09\%$, while the CNN achieves $99.15\%$, reflecting the superior representational capacity of convolutional architectures. Although the CNN exhibits modest robustness gains over the DNN against several attacks, both classical models remain highly vulnerable overall, demonstrating elevated attack success rates across most threat models. In contrast, the HQCNNs demonstrate substantial and consistent robustness improvements. For nearly all attacks, the top-performing HQCNN variant outperforms both classical baselines, with DeepFool being the sole notable exception. Among the HQCNNs, the fully entangled configuration yields the lowest ASR in four attacks, the star-entangled variant in two, and the linear-entangled variant in one. This indicates that the specific entanglement structure plays a meaningful role in shaping adversarial resilience.

\paragraph{Attack-specific observations}
Against classical models, optimization-based attacks such as APGD and C\&W yield the highest attack success rates, underscoring their efficacy against classical architectures. Hybrid models significantly mitigate these vulnerabilities: relative to the CNN baseline, the HQCNN-Linear model reduces the ASR under APGD by $17.44\%$, while the HQCNN-Star model achieves an $89.12\%$ reduction under the C\&W attack. The One-Pixel attack remains largely ineffective across all models, reflecting the inherent stability of MNIST to sparse perturbations. Conversely, the Square Attack proves highly effective against classical models but is dramatically weakened by the HQCNN variants, suggesting enhanced robustness against query-based and black-box perturbations.

\begin{figure}
  \centering
  \includegraphics[width=0.95\textwidth]{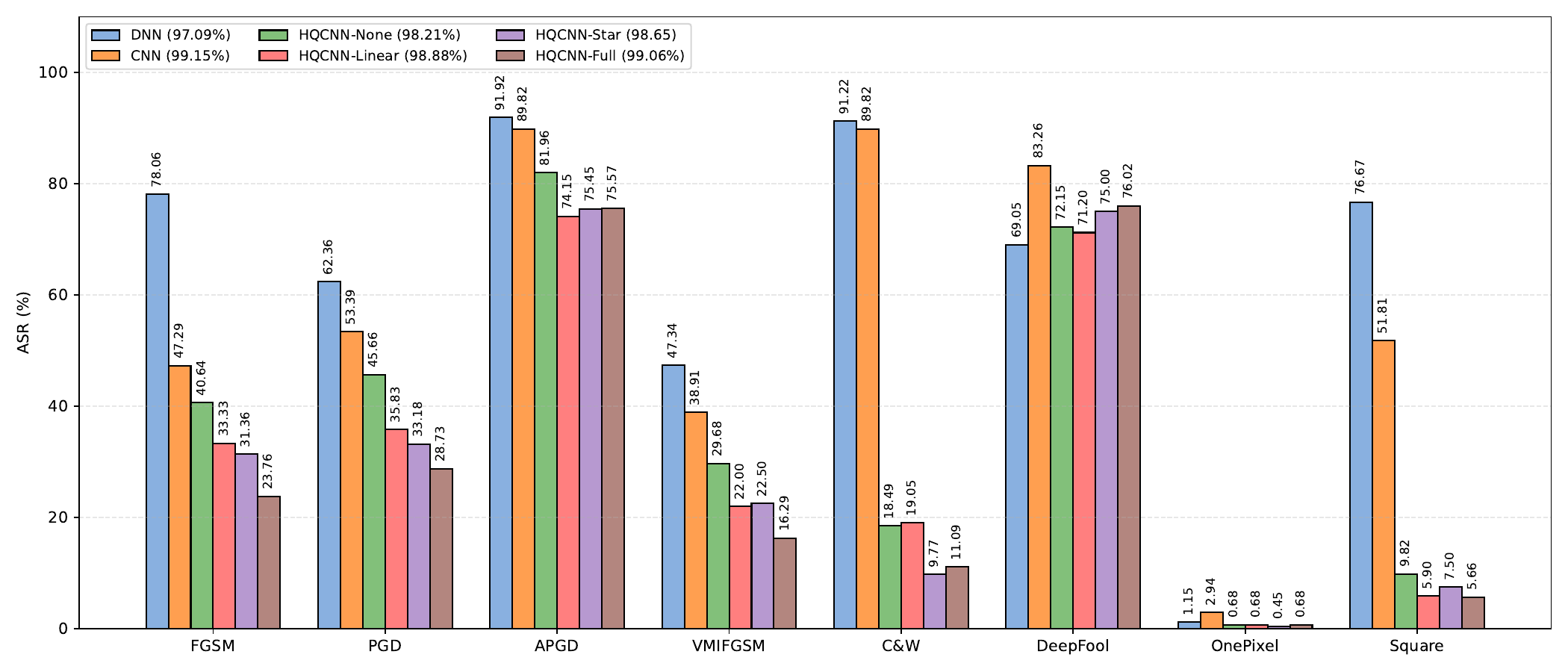}
    \caption{Adversarial attack success rates (ASR) on MNIST. 
    Measured ASR (\%) for DNN, CNN, and HQCNN variants under diverse adversarial attacks, including FGSM, PGD, APGD, VMI-FGSM, C\&W, DeepFool, One-Pixel, and Square Attack. 
    While classical models (DNN, CNN) exhibit higher vulnerability across most attacks, HQCNNs, particularly those with entanglement, achieve substantially lower ASRs, demonstrating improved robustness.}
  \label{fig:mnist_asr_results}
\end{figure}

\paragraph{Summary}
Overall, classical models evaluated on MNIST combine high clean accuracy with pronounced susceptibility to adversarial manipulation. HQCNN models, particularly those incorporating entanglement, offer a highly favorable accuracy--robustness trade-off, maintaining high ODRs (up to $99.06\%$) while substantially reducing the ASR across the majority of attacks. Relative robustness gains range from $17.44\%$ under APGD to $89.12\%$ under C\&W when compared to the CNN baseline. 
Attack success rates for all models on the MNIST dataset are detailed in Fig.~\ref{fig:mnist_asr_results}.

\subsubsection{OrganAMNIST Dataset}
\paragraph{Model-specific observations}
On the OrganAMNIST dataset, the DNN baseline achieves a low ODR of $63.89\%$, indicating limited representational capacity for this medical imaging task. Owing to this poor clean accuracy, subsequent comparisons focus primarily on the CNN as the standard classical baseline. The CNN substantially improves clean performance, reaching an ODR of $89.20\%$, but remains highly susceptible to adversarial perturbations. The HQCNN models exhibit a markedly different robustness profile. Across all attacks, the best-performing HQCNN variant consistently outperforms the CNN baseline in terms of attack success rate. Notably, the HQCNN-Linear model achieves the lowest ASR in seven of the eight evaluated attacks, while the HQCNN-Full model provides the strongest defense in the remaining case, further highlighting the influence of entanglement structure on robustness.

\paragraph{Attack-specific observations}
Optimization-based attacks such as C\&W and DeepFool are particularly effective against classical models, yielding ASRs exceeding $94\%$ for the CNN. In contrast, hybrid models substantially mitigate these vulnerabilities, reducing the ASR by up to $39.04\%$ for the C\&W attack and $7.88\%$ for DeepFool when compared to the CNN baseline. DeepFool remains the most damaging attack overall, inducing high ASRs across both paradigms; however, HQCNN models consistently limit this degradation to the $88$--$93\%$ range, compared to the near-total failure observed in the CNN. The One-Pixel attack proves largely ineffective across all models, with ASRs remaining below $6\%$, suggesting that sparse perturbations are insufficient to reliably mislead classifiers in this medical imaging context. In contrast, the Square Attack reveals one of the sharpest distinctions between paradigms: while classical models exhibit ASRs in the $36$--$40\%$ range, hybrid models suppress these rates to below $8\%$, demonstrating strong robustness against query-based black-box attacks.

\begin{figure}
  \centering
  \includegraphics[width=0.95\textwidth]{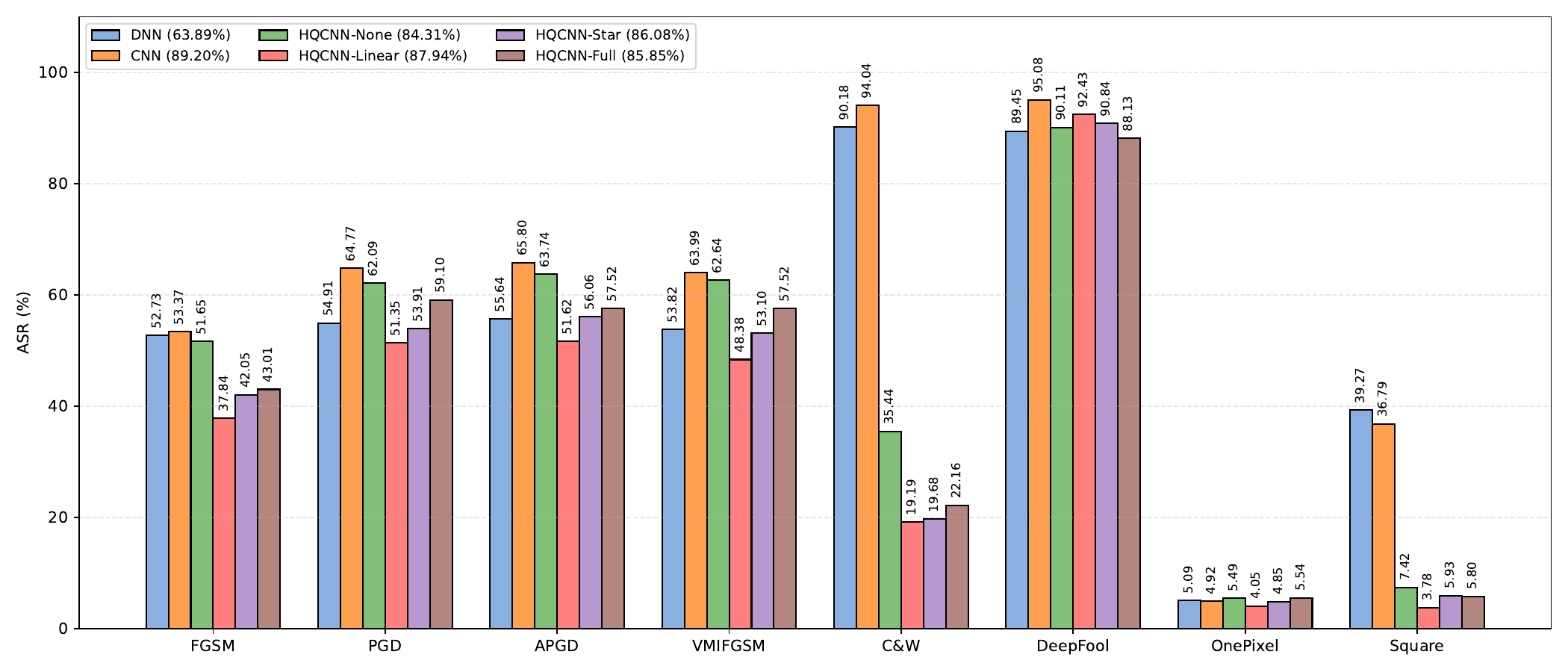}
    \caption{Adversarial attack success rates (ASR) on OrganAMNIST. 
    Measured ASR (\%) for DNN, CNN, and HQCNN variants under a range of adversarial attacks, including FGSM, PGD, APGD, VMI-FGSM, C\&W, DeepFool, One-Pixel, and Square Attack. 
    While the CNN exhibits the highest vulnerability across most attacks, HQCNNs, particularly those with entanglement, reduce ASR and demonstrate improved robustness.}
  \label{fig:organamnist_asr_results}
\end{figure}

\paragraph{Summary}
Overall, although the CNN achieves relatively high clean accuracy on OrganAMNIST, its performance collapses under adversarial attacks. HQCNN models, particularly those incorporating entanglement, achieve a substantially more favorable accuracy--robustness trade-off, maintaining comparable ODRs while dramatically reducing attack success rates. Relative robustness improvements range from $7.30\%$ under the DeepFool attack (HQCNN-Full) to $89.72\%$ under the Square Attack (HQCNN-Linear), compared to the CNN baseline. 
Attack success rates for all evaluated models on the OrganAMNIST dataset are presented in Fig.~\ref{fig:organamnist_asr_results}.

\subsubsection{CIFAR-10 Dataset}
\paragraph{Model-specific observations}
On the CIFAR-10 dataset, the DNN baseline performs poorly, achieving an ODR of only $48.30\%$. Consequently, the CNN model, which achieves a substantially higher ODR of $79.70\%$, is adopted as the primary classical baseline. Despite this strong clean accuracy, the CNN remains highly vulnerable to adversarial perturbations. In contrast, HQCNN models consistently outperform the CNN in terms of robustness across all evaluated attacks. Among these, the HQCNN-Linear variant achieves the lowest attack success rates in five of the eight attacks, while the HQCNN-Full model provides the strongest defense in the remaining three, underscoring the impact of entanglement structure on robustness.

\paragraph{Attack-specific observations}
Optimization-based attacks, particularly C\&W and DeepFool, prove to be the most damaging across all model classes. The CNN baseline experiences near-complete failure under these attacks, with ASRs exceeding $96\%$. HQCNN models substantially reduce, but do not fully eliminate, this vulnerability, lowering ASRs to the $67$--$77\%$ range. The One-Pixel attack is moderately effective against the CNN (ASR of $25.62\%$), while all HQCNN variants constrain its impact to below $22\%$. The most pronounced separation between classical and quantum-enhanced models arises under the Square Attack. While classical models exhibit ASRs in the $19$--$31\%$ range, hybrid models nearly suppress this attack entirely, reducing ASRs to approximately $1$--$3\%$. This behavior highlights the enhanced resistance of HQCNNs to query-based, black-box adversarial strategies.

\begin{figure}
  \centering
  \includegraphics[width=0.95\textwidth]{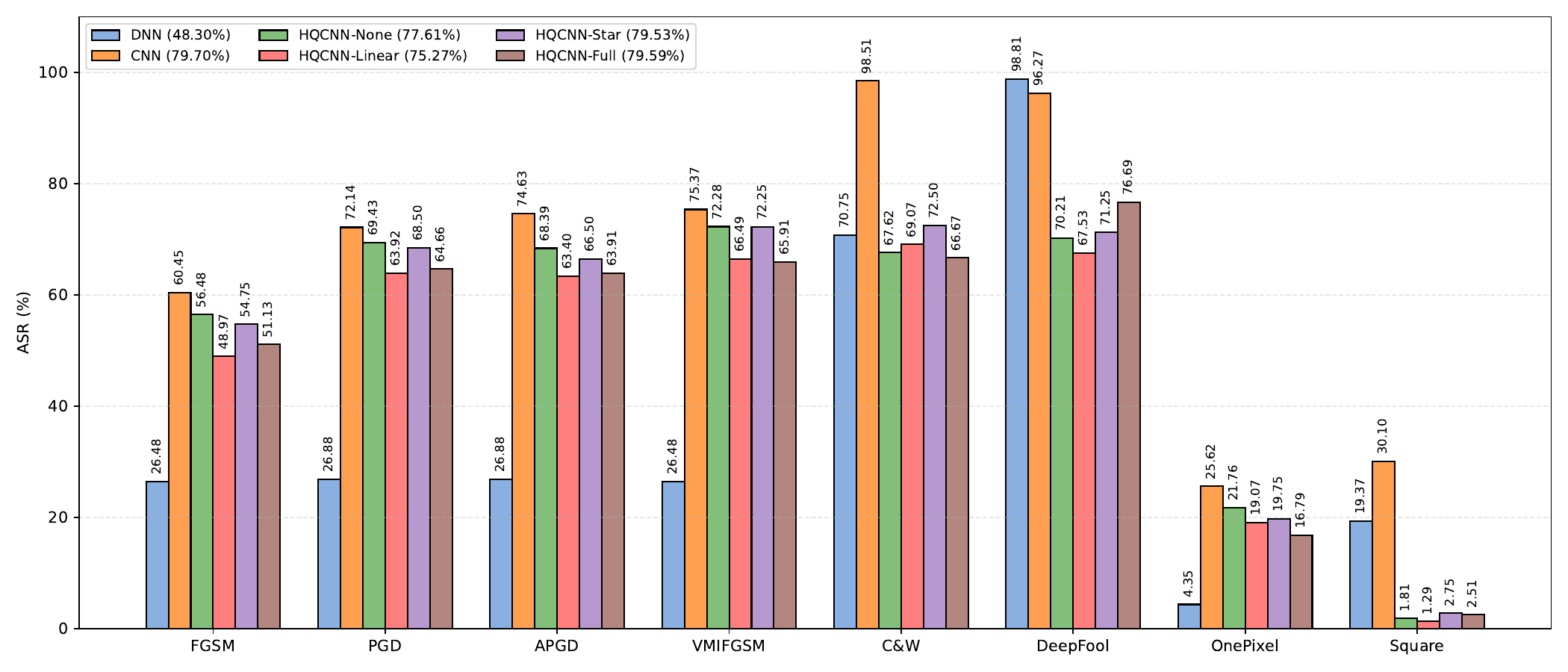}
    \caption{Adversarial attack success rates (ASR) on CIFAR-10. 
    Measured ASR (\%) for DNN, CNN, and HQCNN variants under adversarial attacks, including FGSM, PGD, APGD, VMI-FGSM, C\&W, DeepFool, One-Pixel, and Square Attack. 
    The CNN shows the highest vulnerability across most attacks, while HQCNNs, particularly those with entanglement, achieve lower ASRs, highlighting their enhanced robustness.}
  \label{fig:cifar10_asr_results}
\end{figure}

\paragraph{Summary}
Overall, classical models evaluated on the CIFAR-10 dataset exhibit a clear accuracy--robustness trade-off: the DNN lacks sufficient expressive power, whereas the CNN achieves higher clean accuracy but collapses under adversarial attacks. HQCNN models strike a more favorable balance, maintaining competitive ODRs while significantly improving robustness. Relative robustness gains range from $11.39\%$ under PGD to an exceptional $95.71\%$ under the Square Attack, both achieved by the HQCNN-Linear model. While HQCNN-Linear offers the strongest overall robustness, the HQCNN-Full model preserves both high accuracy ($79.59\%$) and improved robustness. 
Fig.~\ref{fig:cifar10_asr_results} illustrates the attack success rates for all evaluated models on the CIFAR-10 dataset.

\subsection{Runtime Analysis of Adversarial Attacks}
To complement the robustness evaluation, we analyze the computational cost required to generate adversarial examples across all attacks and datasets. Runtime is reported in seconds on a logarithmic scale to accommodate the wide variability in attack complexity and execution times. Importantly, adversarial examples are generated exclusively on the subset of test samples that are correctly classified by the target model prior to the attack. This ensures that runtime measurements reflect the true cost of inducing misclassification, rather than trivially including already incorrect predictions. This analysis highlights the trade-off between adversarial robustness and the computational effort required to mount effective attacks.

Across all three datasets, consistent trends emerge. Gradient-based attacks (e.g., FGSM and PGD) are the most efficient, completing within approximately $1$--$11\,\mathrm{seconds}$ on the DNN and $3$--$95\,\mathrm{seconds}$ on the CNN. In contrast, optimization-based attacks such as C\&W and DeepFool are substantially more expensive, requiring up to $388\,\mathrm{seconds}$ on the DNN and exceeding $2{,}250\,\mathrm{seconds}$ on the CNN. Query-based attacks (One-Pixel and Square Attack) also incur high computational costs, with runtimes surpassing $52\,\mathrm{seconds}$ on the DNN and $345\,\mathrm{seconds}$ on the CNN.

The HQCNN models impose a markedly higher computational burden on adversarial generation. Even the single-step FGSM attack exceeds $148\,\mathrm{seconds}$, while iterative and query-based attacks range from approximately $1{,}373\,\mathrm{seconds}$ (PGD) to over $85{,}000\,\mathrm{seconds}$ (Square Attack). Moreover, runtime scales systematically with entanglement complexity: models with greater entanglement incur higher attack-generation costs, with the HQCNN-Full variant being the most computationally demanding across nearly all attack scenarios and datasets. Several clear patterns emerge from this evaluation:

\begin{itemize}
    \item \textbf{Attack-type Dependence.} FGSM remains the fastest attack across all models but becomes orders of magnitude slower on hybrid models. Iterative methods scale with the number of optimization steps, while optimization- and query-based attacks dominate overall runtime.
    \item \textbf{Entanglement Overhead.} Increased entanglement consistently amplifies attack-generation costs, with fully entangled HQCNNs exhibiting the highest runtimes across nearly all attack scenarios.
\end{itemize}

It is critical to emphasize that absolute runtime is an inherently imperfect metric for comparing classical and quantum-enhanced models. Because current HQCNN runtimes are based on classical simulations, where repeated circuit evaluations constitute the primary bottleneck, these figures do not reflect future quantum hardware performance, energy usage, or operational costs. Consequently, these runtimes should be viewed as a practical proxy for current attack difficulty rather than a definitive measure of real-world deployment costs.

Within this context, the results reveal a fundamental trade-off: classical models are comparatively inexpensive to attack, whereas hybrid models, particularly those with complex entanglement, achieve higher robustness while simultaneously imposing a substantial computational barrier to adversarial example generation. This suggests that quantum-enhanced architectures may confer robustness not only through improved decision boundaries but also by increasing the computational cost of adversarial optimization.

\begin{figure}
  \centering
  \includegraphics[width=0.95\textwidth]{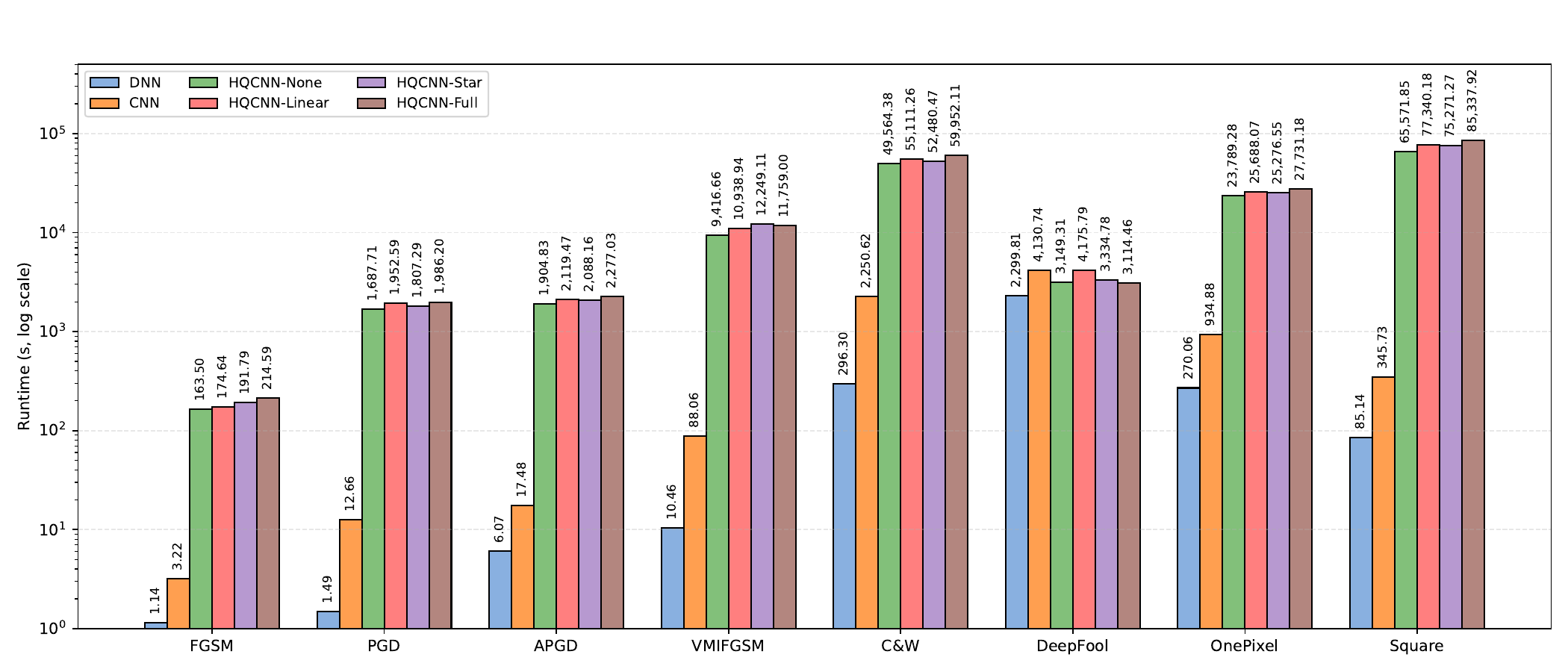}
    \caption{Adversarial attack runtimes on MNIST. 
    Measured adversarial example generation times (seconds, log scale) for FGSM, PGD, APGD, VMI-FGSM, C\&W, DeepFool, One-Pixel, and Square attacks across DNN, CNN, and HQCNN variants. 
    Adversarial attacks on classical models (DNN, CNN) achieve much faster runtimes, whereas HQCNNs incur significantly higher computational costs, with runtimes increasing alongside entanglement complexity.}
  \label{fig:mnist_attack_runtime}
\end{figure}

\begin{figure}
  \centering
  \includegraphics[width=0.95\textwidth]{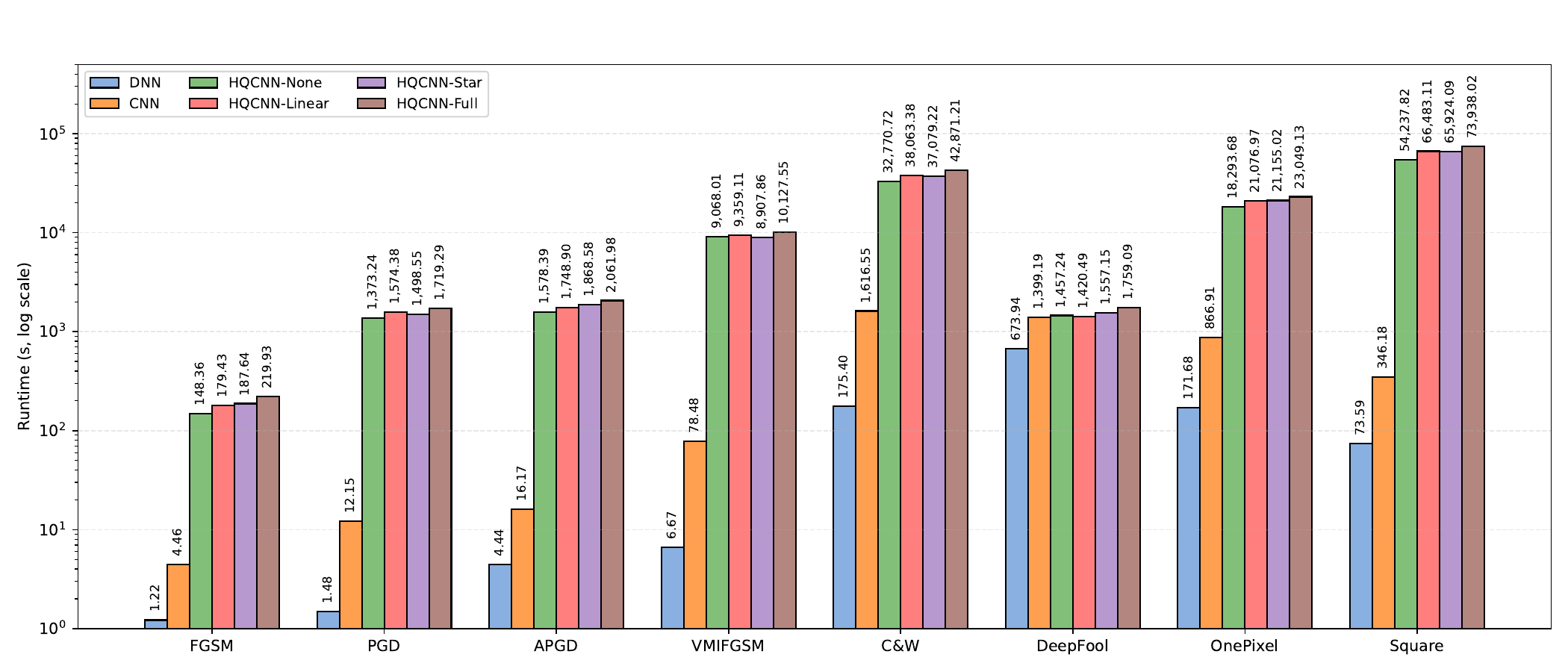}
    \caption{Adversarial attack runtimes on OrganAMNIST. 
    Measured adversarial example generation times (seconds, log scale) for FGSM, PGD, APGD, VMI-FGSM, C\&W, DeepFool, One-Pixel, and Square attacks across DNN, CNN, and HQCNN variants. 
    Adversarial attacks on classical models (DNN, CNN) achieve much faster runtimes, whereas HQCNNs incur significantly higher computational costs, with runtimes increasing alongside entanglement complexity.}
  \label{fig:organamnist_attack_runtime}
\end{figure}

\begin{figure}
  \centering
  \includegraphics[width=0.95\textwidth]{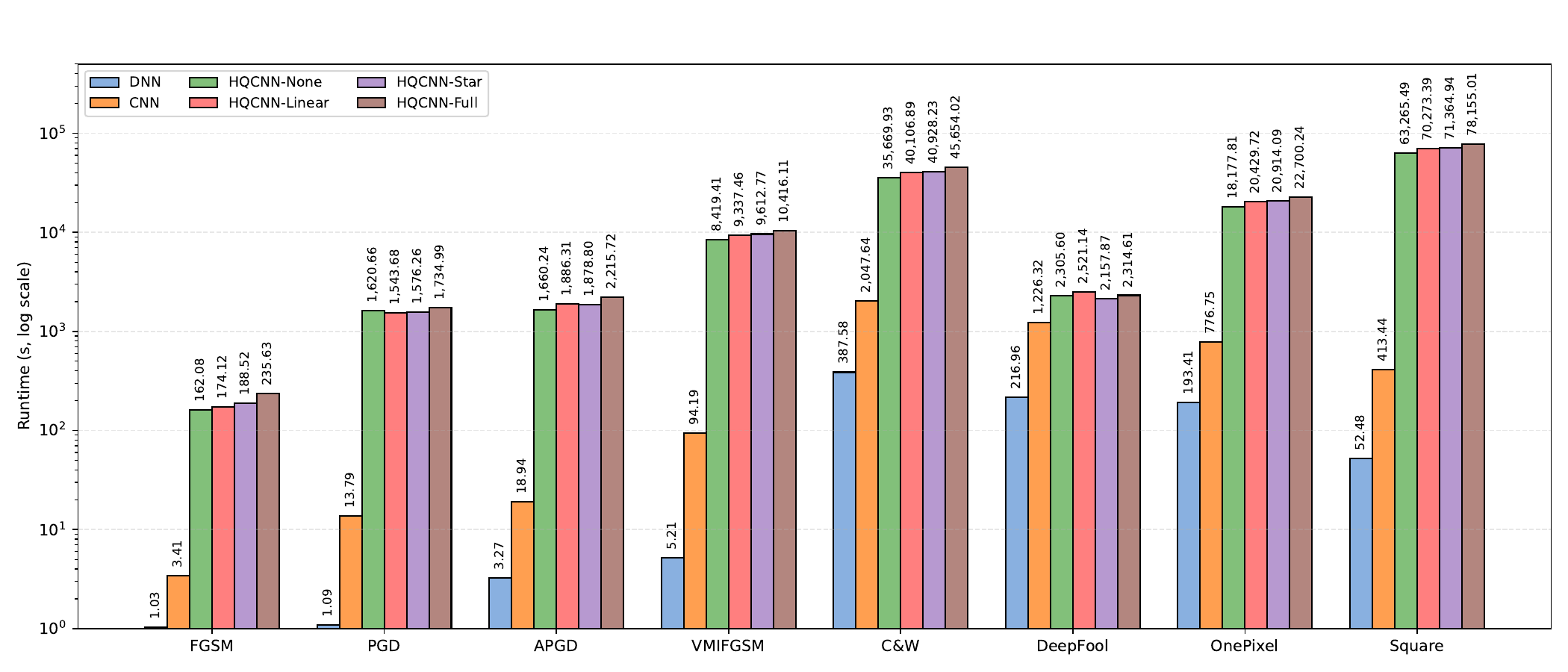}
    \caption{Adversarial attack runtimes on CIFAR-10. 
    Measured adversarial example generation times (seconds, log scale) for FGSM, PGD, APGD, VMI-FGSM, C\&W, DeepFool, One-Pixel, and Square attacks across DNN, CNN, and HQCNN variants. 
    Adversarial attacks on classical models (DNN, CNN) achieve much faster runtimes, whereas HQCNNs incur significantly higher computational costs, with runtimes increasing alongside entanglement complexity.}
  \label{fig:cifar10_attack_runtime}
\end{figure}

Runtime measurements for MNIST, OrganAMNIST, and CIFAR-10 datasets are visualized in Figs.~\ref{fig:mnist_attack_runtime}, \ref{fig:organamnist_attack_runtime}, and \ref{fig:cifar10_attack_runtime}, respectively.

\FloatBarrier

\section{Conclusion}
In this work, we systematically evaluated the adversarial robustness of classical neural networks alongside our proposed modular hybrid quantum--classical architectures across the MNIST, OrganAMNIST, and CIFAR-10 datasets. While the classical CNN consistently achieved the highest clean accuracy, it exhibited marked vulnerability to adversarial perturbations, frequently attaining high Original Detection Rates while simultaneously suffering from high Attack Success Rates. This behavior underscores a clear accuracy--robustness trade-off inherent in conventional architectures.

In contrast, the proposed hybrid quantum--classical neural networks, particularly those incorporating entanglement structures, demonstrated a significantly more favorable balance between accuracy and robustness. Across all evaluated datasets, entangled HQCNN variants preserved competitive ODRs while substantially reducing ASRs, indicating enhanced resistance to adversarial attacks. On MNIST, relative robustness gains ranged from $17.44\%$ under APGD to $89.12\%$ under the C\&W attack. Similar trends were observed on OrganAMNIST, with improvements spanning $7.30\%$ under DeepFool to $89.72\%$ under the Square Attack, and on CIFAR-10, where gains ranged from $11.39\%$ under PGD to an exceptional $95.71\%$ under the Square Attack.

The adversarial runtime analysis revealed a complementary dimension of robustness. Classical models are comparatively inexpensive to attack, rendering them practically more vulnerable in real-world scenarios. Hybrid models, by contrast, impose a substantial computational barrier on adversarial example generation. Increased entanglement complexity directly correlates with higher attack-generation costs, with the fully entangled HQCNN often exhibiting the greatest runtime overhead while concurrently delivering strong robustness. These results highlight a fundamental trade-off: hybrid quantum--classical models demand greater computational resources but significantly elevate the operational cost of executing successful adversarial attacks.

Overall, the proposed QShield architecture significantly enhances adversarial robustness while maintaining strong clean-data performance. By integrating a classical CNN backbone with a noise-aware, entanglement-based quantum module and an adaptive fusion mechanism, QShield achieves a hybrid model that is both highly expressive and resilient.

Taken together, these findings position QShield as a highly promising architecture for secure and reliable machine learning, particularly in sensitive domains such as medical imaging and safety-critical computer vision, where robustness against adversarial attacks is essential.


\backmatter

\section*{Declarations}

\bmhead{Funding}
The authors received no financial support from any organization for the research, authorship, and/or publication of this article.

\bmhead{Conflict of Interest}
The authors declare that they have no relevant financial or non-financial interests to disclose.

\bmhead{Code Availability}
The source code, datasets, and pretrained models supporting the findings of this work are publicly available in the following GitHub repository: \url{https://github.com/n-azimi/QShield}.


\begin{appendices}
\section{Adversarial Attacks: Hyperparameters}
\label{appendix:attack_params}
This appendix documents the comprehensive set of hyperparameter configurations utilized for all adversarial attacks throughout our experimental evaluation on the MNIST, OrganAMNIST, and CIFAR-10 datasets. The reported parameters adhere to widely adopted conventions within the adversarial machine learning literature. They were carefully selected to guarantee strong attack efficacy while preserving consistency across the various datasets and model architectures.

Table~\ref{tab:adversarial_attack_parameters} presents a detailed summary of the attack-specific configurations, including perturbation budgets, step sizes, iteration counts, and optimizer settings. Unless explicitly stated otherwise, all attacks are untargeted and evaluated under identical implementation conditions to facilitate reproducibility and ensure fair comparisons.

\begin{table}[htbp]
\centering
\caption{Hyperparameter settings for adversarial attacks across all experimental datasets.}
\vspace{0.3cm}
\label{tab:adversarial_attack_parameters}

\footnotesize
\begin{tabular}{@{}llp{8cm}@{}}
\toprule
\textbf{Dataset} & \textbf{Attack Method} & \textbf{Hyperparameter Configuration} \\
\midrule
\multirow{13}{*}{\rotatebox{90}{\hspace{10pt}\textbf{MNIST}}}
& FGSM & $\epsilon = 32/255$ \\
\cmidrule{2-3}
& PGD & $\epsilon = 32/255$, step size $\alpha = 2/255$, steps = 10, random initialization = True \\
\cmidrule{2-3}
& APGD & $\ell_\infty$ norm, $\epsilon = 32/255$, steps = 10, restarts = 1, seed = 0, loss = cross-entropy, EOT iterations = 1, step size update factor $\rho = 0.75$ \\
\cmidrule{2-3}
& VMI-FGSM & $\epsilon = 32/255$, step size $\alpha = 2/255$, steps = 10, momentum decay = 1.0, sampled examples in the neighborhood $N = 5$, neighborhood upper bound $\beta = 1.5$ \\
\cmidrule{2-3}
& C\&W & $\ell_2$ norm, confidence = 0.0, untargeted attack, learning rate = 0.05, binary search steps = 10, maximum iterations = 5, initial constant = 0.01, halving/doubling limits = 5, batch size = 128 \\
\cmidrule{2-3}
& DeepFool & steps = 50, overshoot parameter = 0.05 \\
\cmidrule{2-3}
& One-Pixel & pixels = 1, steps = 10, population size = 10, inference batch size = 128 \\
\cmidrule{2-3}
& Square Attack & $\ell_\infty$ norm, $\epsilon = 32/255$, queries = 500, restarts = 1, control size of squares parameter $p_{init} = 0.8$, margin loss, rescaling schedule = True, seed = 0 \\
\midrule
\multirow{13}{*}{\rotatebox{90}{\hspace{10pt}\textbf{OrganAMNIST}}}
& FGSM & $\epsilon = 8/255$ \\
\cmidrule{2-3}
& PGD & $\epsilon = 8/255$, step size $\alpha = 2/255$, steps = 10, random initialization = True \\
\cmidrule{2-3}
& APGD & $\ell_\infty$ norm, $\epsilon = 8/255$, steps = 10, restarts = 1, seed = 0, loss = cross-entropy, EOT iterations = 1, step size update factor $\rho = 0.75$ \\
\cmidrule{2-3}
& VMI-FGSM & $\epsilon = 8/255$, step size $\alpha = 2/255$, steps = 10, momentum decay = 1.0, sampled examples in the neighborhood $N = 5$, neighborhood upper bound $\beta = 1.5$ \\
\cmidrule{2-3}
& C\&W & $\ell_2$ norm, confidence = 0.0, untargeted attack, learning rate = 0.05, binary search steps = 8, maximum iterations = 5, initial constant = 0.01, halving/doubling limits = 5, batch size = 128 \\
\cmidrule{2-3}
& DeepFool & steps = 50, overshoot parameter = 0.05 \\
\cmidrule{2-3}
& One-Pixel & pixels = 1, steps = 10, population size = 10, inference batch size = 128 \\
\cmidrule{2-3}
& Square Attack & $\ell_\infty$ norm, $\epsilon = 8/255$, queries = 500, restarts = 1, control size of squares parameter $p_{init} = 0.8$, margin loss, rescaling schedule = True, seed = 0 \\
\midrule
\multirow{13}{*}{\rotatebox{90}{\hspace{10pt}\textbf{CIFAR-10}}}
& FGSM & $\epsilon = 2/255$ \\
\cmidrule{2-3}
& PGD & $\epsilon = 2/255$, step size $\alpha = 2/255$, steps = 10, random initialization = True \\
\cmidrule{2-3}
& APGD & $\ell_\infty$ norm, $\epsilon = 2/255$, steps = 10, restarts = 1, seed = 0, loss = cross-entropy, EOT iterations = 1, step size update factor $\rho = 0.75$ \\
\cmidrule{2-3}
& VMI-FGSM & $\epsilon = 2/255$, step size $\alpha = 2/255$, steps = 10, momentum decay = 1.0, sampled examples in the neighborhood $N = 5$, neighborhood upper bound $\beta = 1.5$ \\
\cmidrule{2-3}
& C\&W & $\ell_2$ norm, confidence = 0.0, untargeted attack, learning rate = 0.05, binary search steps = 8, maximum iterations = 5, initial constant = 0.01, halving/doubling limits = 5, batch size = 128 \\
\cmidrule{2-3}
& DeepFool & steps = 50, overshoot parameter = 0.05 \\
\cmidrule{2-3}
& One-Pixel & pixels = 1, steps = 10, population size = 10, inference batch size = 128 \\
\cmidrule{2-3}
& Square Attack & $\ell_\infty$ norm, $\epsilon = 2/255$, queries = 500, restarts = 1, control size of squares parameter $p_{init} = 0.8$, margin loss, rescaling schedule = True, seed = 0 \\
\botrule
\end{tabular}
\end{table}


\section{Adversarial Attacks: Runtime}
\label{appendix:attack_runtime}

This appendix details the computational runtime associated with adversarial example generation across the various attack methodologies and model architectures. Specifically, we measure the average wall-clock time required to generate a single adversarial sample for each attack on the MNIST, OrganAMNIST, and CIFAR-10 datasets.

Table~\ref{tab:attack_runtimes} presents the per-sample runtime results for the classical DNN and CNN baselines, alongside the proposed HQCNN models under their respective entanglement configurations (None, Linear, Star, and Full). All reported measurements are averaged over the entire test sets and were obtained under strictly identical hardware and software environments to guarantee consistency and comparability.

\begin{table}[htbp]
\centering
\caption{Average wall-clock time (in seconds) to generate a single adversarial example across datasets.}
\vspace{0.3cm}
\label{tab:attack_runtimes}

\footnotesize
\begin{tabular*}{\textwidth}{@{\extracolsep\fill}lccccccc@{}}
\toprule
\multirow{3.5}{*}{\textbf{Dataset}} & \multirow{3.5}{*}{\textbf{Attack}} & \multicolumn{6}{c}{\textbf{Average Per-Sample Adversarial Runtime (s)}} \\
\cmidrule{3-8}
 & & \multirow{2.5}{*}{\textbf{DNN}} & \multirow{2.5}{*}{\textbf{CNN}} & \textbf{HQCNN-} & \textbf{HQCNN-} & \textbf{HQCNN-} & \textbf{HQCNN-} \\
 & & & & \textbf{None} & \textbf{Linear} & \textbf{Star} & \textbf{Full} \\
\midrule
\multirow{8}{*}{\rotatebox{90}{\textbf{MNIST}}} 
& FGSM & 0.00013 & 0.00036 & 0.01866 & 0.01980 & 0.02179 & 0.02427 \\ \cmidrule{2-8}
& PGD & 0.00017 & 0.00143 & 0.19266 & 0.22138 & 0.20537 & 0.22468 \\ \cmidrule{2-8}
& APGD & 0.00070 & 0.00197 & 0.21744 & 0.24030 & 0.23729 & 0.25758 \\ \cmidrule{2-8}
& VMI-FGSM & 0.00120 & 0.00996 & 1.07496 & 1.24024 & 1.39194 & 1.33020 \\ \cmidrule{2-8}
& C\&W & 0.03421 & 0.25459 & 5.65803 & 6.24844 & 5.96368 & 6.78191 \\ \cmidrule{2-8}
& DeepFool & 0.26556 & 0.46727 & 0.35951 & 0.47344 & 0.37895 & 0.35231 \\ \cmidrule{2-8}
& One-Pixel & 0.03118 & 0.10575 & 2.71567 & 2.91247 & 2.87233 & 3.13701 \\ \cmidrule{2-8}
& Square & 0.00983 & 0.03910 & 7.48537 & 8.76872 & 8.55355 & 9.65361 \\
\midrule
\multirow{8}{*}{\rotatebox{90}{\textbf{OrganAMNIST}}} 
& FGSM & 0.00022 & 0.00057 & 0.02037 & 0.02424 & 0.02528 & 0.02901 \\ \cmidrule{2-8}
& PGD & 0.00026 & 0.00157 & 0.18863 & 0.21275 & 0.20196 & 0.22681 \\ \cmidrule{2-8}
& APGD & 0.00080 & 0.00209 & 0.21681 & 0.23633 & 0.25183 & 0.27202 \\ \cmidrule{2-8}
& VMI-FGSM & 0.00121 & 0.01016 & 1.24560 & 1.26474 & 1.20052 & 1.33608 \\ \cmidrule{2-8}
& C\&W & 0.03189 & 0.20939 & 4.50147 & 5.14370 & 4.99719 & 5.65583 \\ \cmidrule{2-8}
& DeepFool & 0.12253 & 0.18124 & 0.20017 & 0.19195 & 0.20985 & 0.23206 \\ \cmidrule{2-8}
& One-Pixel & 0.03121 & 0.11229 & 2.51286 & 2.84823 & 2.85108 & 3.04078 \\ \cmidrule{2-8}
& Square & 0.01338 & 0.04484 & 7.45025 & 8.98420 & 8.88464 & 9.75435 \\
\midrule
\multirow{8}{*}{\rotatebox{90}{\textbf{CIFAR-10}}} 
& FGSM & 0.00020 & 0.00042 & 0.02099 & 0.02243 & 0.02356 & 0.02952 \\ \cmidrule{2-8}
& PGD & 0.00021 & 0.00171 & 0.20993 & 0.19892 & 0.19703 & 0.21741 \\ \cmidrule{2-8}
& APGD & 0.00064 & 0.00235 & 0.21505 & 0.24308 & 0.23485 & 0.27765 \\ \cmidrule{2-8}
& VMI-FGSM & 0.00102 & 0.01171 & 1.09059 & 1.20328 & 1.20159 & 1.30527 \\ \cmidrule{2-8}
& C\&W & 0.07659 & 0.25468 & 4.62045 & 5.16841 & 5.11602 & 5.72105 \\ \cmidrule{2-8}
& DeepFool & 0.04287 & 0.15252 & 0.29865 & 0.32488 & 0.26973 & 0.29005 \\ \cmidrule{2-8}
& One-Pixel & 0.03822 & 0.09661 & 2.35463 & 2.63269 & 2.61426 & 2.84464 \\ \cmidrule{2-8}
& Square & 0.01037 & 0.05142 & 8.19501 & 9.05584 & 8.92061 & 9.79386 \\
\botrule
\end{tabular*}
\end{table}


\end{appendices}

\bibliography{ref}

\end{document}